\documentclass[12pt]{article}

\usepackage[utf8]{inputenc}
\usepackage{cancel}
\usepackage{amssymb}
\usepackage{soul}
\usepackage{tabularx}
\usepackage{xcolor}
\usepackage{booktabs}
\usepackage{subcaption} 
\usepackage{colortbl}
\usepackage{authblk}
\DeclareUnicodeCharacter{00A0}{ }

\newcolumntype{C}{>{\centering\arraybackslash}X}
\usepackage[T1]{fontenc}
\usepackage{epsfig}
\usepackage{latexsym}
\usepackage{graphicx}
\usepackage{amsmath}
\usepackage{amsfonts}   
\usepackage{amssymb}    
\usepackage{float}
\usepackage{bm}
\usepackage{url}
\usepackage{hyperref} 
\usepackage[nodisplayskipstretch]{setspace}
\def\lsim{\raise0.3ex\hbox{$\;<$\kern-0.75em\raise-1.1ex\hbox{$\sim\;$}}}
\def\gsim{\raise0.3ex\hbox{$\;>$\kern-0.75em\raise-1.1ex\hbox{$\sim\;$}}}

\newcommand{\captions}{\sf\caption}
\def    \beq            {\begin{equation}}
\def    \eeq            {\end{equation}}
\def    \bea           {\begin{eqnarray}}
\def    \eea           {\end{eqnarray}}

\def\g2{{\rm GeV}^2}

\def\sw2{sin^2 \theta_w}

\def\a^tau{\alpha_{\tau}}

\def\beq{\begin{equation}}
\def\eeq{\end{equation}}
\def\beqa{\begin{eqnarray}}
\def\eeqa{\end{eqnarray}}

\newcommand{\newc}{\newcommand}
\newc\BR{BR}
\newc{\akappa}{A_{\kappa} }
\newc\deltagmtwo{\delta (g-2)_{\mu}} 
\newc\deltaamu{\Delta a_{\mu}}

\def\anti{\overline}

\newc{\haa}{BR\(h_1\to a_1 a_1\)}
\newc{\abb}{BR\(a_1\to b\anti{b}\)}
\newc{\hbb}{BR\(h_1\to b\anti{b}\)}
\newc{\abund}{\Omega h^2}
\newc\bsgamma{b\rightarrow s \gamma }
\newc\bxsgamma{\overline{B}\rightarrow X_{s}\gamma}
\newc\brbsgamma{\BR(\overline{B}\rightarrow X_s\gamma)}

\newc{\Fermi}{\textit{Fermi}-}

\allowdisplaybreaks

\usepackage{array, makecell}
\usepackage{boldline}
\usepackage{cite}

\title{\bf{
Right-handed sneutrino and gravitino multicomponent dark matter in light of neutrino detectors
}}
\author[a]{Jong~Soo~Kim\thanks{jongsoo.kim@tu-dortmund.de}}
\author[b,c]{Daniel~E.~López-Fogliani\thanks{daniel.lopez@df.uba.ar}}
\author[d]{Andres~D.~Perez\thanks{andres.perez@iflp.unlp.edu.ar}}
\author[e]{Roberto~Ruiz~de~Austri\thanks{rruiz@ific.uv.es}}
\affil[a]{School of Physics, University of the Witwatersrand, Johannesburg, South Africa}
\affil[b]{Instituto de Física de Buenos Aires UBA \& CONICET, Departamento de Física,
Facultad de Ciencia Exactas y Naturales, Universidad de Buenos Aires, 
1428 Buenos Aires, Argentina}
\affil[c]{
{Pontificia Universidad Católica Argentina, 
Av. Alicia Moreau de Justo 1500, 
1107~Buenos~Aires, Argentina}}
\affil[d]{IFLP, CONICET - Dpto. de Física, Universidad Nacional de La Plata,\protect\\
C.C. 67, 1900 La Plata, Argentina}
\affil[e]{Instituto de Física Corpuscular CSIC-UV, c/Catedrático José Beltrán 2, 46980 Paterna (Valencia), Spain}
\renewcommand\Affilfont{\itshape\small}

\date{}

\begin{document}
\maketitle
\begin{abstract}

We investigate the possibility that right-handed (RH) sneutrinos and gravitinos can coexist and explain the dark matter (DM) problem. We compare extensions of the minimal supersymmetric standard model (MSSM) and the next-to-MSSM (NMSSM) adding RH neutrinos superfields, with special emphasis on the latter. If the gravitino is the lightest supersymmetric particle (LSP) and the RH sneutrino the next-to-LSP (NLSP), the heavier particle decays to the former plus left-handed (LH) neutrinos through the mixing between the scalar partners of the LH and RH neutrinos.
However, the interaction is suppressed by the Planck mass, and if the LH-RH sneutrino mixing parameter is small, $\ll O(10^{-2})$, a long-lived RH sneutrino NLSP is possible even surpassing the age of the Universe.
As a byproduct, the NLSP to LSP decay produces monochromatic neutrinos in the ballpark of current and planned neutrino telescopes like Super-Kamiokande, IceCube and Antares that we use to set constraints and show prospects of detection. 
In the NMSSM+RHN, assuming a gluino mass parameter $M_3 = 3$ TeV we found the following lower limits for the gravitino mass $m_{3/2} \gtrsim 1-600$ GeV and the reheating temperature $T_R \gtrsim 10^5 - 3 \times 10^7$ GeV, for $m_{\tilde{\nu}_R} \sim 10-800$ GeV. If we take $M_3=10$ TeV, then the limits on $T_R$ are relaxed by one order of magnitude.

\end{abstract}

{\small   Keywords: Supersymmetry, Dark Matter, Gravitino, Sneutrino.} 

\newpage

\tableofcontents 


\section{Introduction}
\label{sec:intro}

Despite the huge experimental effort of the last decades, the composition of the dark matter (DM) sector in the Universe~\cite{Planck:2018vyg,Bertone09,Feng:2010} remains unknown. Although most works in this field assume that DM is dominated by only one type of particle, an intriguing possibility is that multiple candidates can coexist and contribute significantly to the measured relic abundance. This hypothesis has also been studied, see for example Refs.~\cite{Zurek:2008qg,Profumo:2009tb,Arcadi:2016kmk,Bhattacharya:2018cgx,Chakraborti:2018aae,Bhattacharya:2019tqq,Betancur:2020fdl,Khalil:2020syr,Belanger:2021lwd}, where the DM components are usually stable by a discrete symmetry such as $Z_2 \times Z_2$ or $Z_N$ with $N>4$, or in Refs.~\cite{Baer:2010kd,Bae:2013hma} by including axions as a second stable DM candidate alongside neutralino DM. In the context of $R$-parity breaking models, in Refs.~\cite{Gomez-Vargas:2019vci,Gomez-Vargas:2019mqk} a mixture of gravitinos and axinos was studied.

In this paper we focus on $R$-parity conserving supersymmetric models (SUSY) in which the typical DM candidate would be a WIMP. However if its relic abundance lies below the measured value the following questions arise, can an additional candidate coexist and naturally provide the missing relic amount? And if so, can we look for new signals to distinguish between the vanilla single DM and the multicomponent DM case?
To address these issues we consider a two-component scenario with right-handed (RH) sneutrinos~\footnote{Sneutrinos are the scalar partners of the neutrinos. However, only RH sneutrino can be a good DM candidate, since its LH counterpart can not account for a significant fraction of the DM relic density~\cite{Falk:1994,Arina:2007}} and gravitinos~\footnote{Gravitinos are the supersymmetric partners of graviton particles, and can be valid non-WIMP DM candidates for both $R$-parity conserving~\cite{Takayama:2000uz,Hamaguchi:2017} and $R$-parity breaking models~\cite{Takayama:2000uz,Buchmuller:2007ui,Bertone:2007aw,Ibarra:2007wg,Ishiwata:2008cu,Choi:2009ng,Choi:2010xn,Choi:2010jt,Diaz:2011pc,Restrepo:2011rj,GomezVargas:2017,Gomez-Vargas:2019mqk}.} as DM candidates. We compare extensions of the well-known minimal supersymmetric standard model (MSSM) and the next-to-MSSM (NMSSM) by adding RH neutrino superfields, denoted MSSM+RHN~\cite{Arkani:2001} and NMSSM+RHN~\cite{Kitano:2000}, respectively. An important characteristic is that we do not include any extra symmetry, therefore one of the candidates would not be stable since it must decay to the lightest one (however we will see that it can be stable on cosmological time scales). A similar situation was studied in Ref.~\cite{Essig:2013goa} considering a supersymmetric hidden sector, with an additional $U(1)$ gauge group, although with only one candidate as the dominant contribution of the relic abundance.

In particular, we consider the gravitino as the lightest supersymmetric particle (LSP), and the RH sneutrino as the next-to-LSP (NSLP). 
Since the gravitino LSP is essentially decoupled from the rest of the spectrum, its presence does not affect the evolution of other thermal relics. Therefore to calculate the initial RH sneutrino relic density we can ignore the gravitino in our computations and reintroduce it afterwards. Collider signatures with this mass hierarchy have been studied previously in the context of other models~\cite{Covi:2007,Ellis:2008as}.

Interestingly, the NLSP decay is dominated by two-body processes to the LSP plus neutrinos, producing a potentially detectable signal as an excess over the expected atmospheric-neutrino background measured by neutrino telescopes in the GeV range. In Ref.~\cite{Dhuria:2018}, the authors consider a PeV scale SUSY breaking framework to explain the observation of very high energy neutrino events at IceCube, but with a tiny fraction ($10^{-6}$) of sneutrino DM. In the current analysis, we are presenting a well-known low energy phenomenological SUSY realization where the RH sneutrino NLSP lifetime can be longer than the age of the Universe with a significant contribution to the total DM relic density (a fraction greater than $10^{-2}$). One important condition that we will consider is RH neutrinos heavier than the RH sneutrino NLSP, otherwise the latter would not be a good DM candidate since an extra decay channel would open making its lifetime typically very short. Besides that, the multicomponent DM set-up can be achieve without fine-tuning the DM masses because RH sneutrino-gravitino interactions are suppressed by the Planck mass scale and by the small mixing angle between left-handed (LH) and RH sneutrinos, $\ll O(10^{-2})$, that can be found for typical parameter values in the NMSSM+RHN. As we will see, the mixing angle between LH and RH sneutrinos becomes a crucial parameter to determine the multicomponent scenario phenomenology.

Neutrinos, like photons, are unaffected by magnetic fields and thus allow to reconstruct the direction of their origin. However, to solve the difficulty of large expected atmospheric neutrino background for energies under few TeV, very large detectors are needed. The energy ranges of the Cherenkov detectors Super-Kamiokande~\cite{FukudaSK:2003}, Antares~\cite{ANTARES:2011hfw} and IceCube~\cite{IceCube:2016zyt} make them ideal instruments to constraint the parameter space of decaying RH sneutrinos. Even more, upcoming experiments like the next operational phase of Super-Kamiokande, called Hyper-Kamiokande~\cite{HyperK:2018}, the IceCube Upgrade~\cite{Ishihara:2019aao}, KM3NeT~\cite{KM3Net:2016zxf}, and the liquid argon time-projection chamber (LArTPC) DUNE~\cite{DUNE:2018}, will further explore promising regions of the model in the near future. Then, we study the coexisting RH sneutrino NLSP plus gravitino LSP DM scenario, and the possible neutrino smoking-gun signal with current and future facilities. Recall that besides this extra signature, the RH sneutrino has to satisfy the usual WIMP constraints.

Regarding the theoretical framework, the introduction of RH neutrino superfields in supersymmetric models with $R$-parity conservation not only provides an interesting extra DM candidate, the aforementioned sneutrino, but allows to reproduce the measured neutrino pattern~\cite{Pontecorvo:1969,FukudaSK:1998,Ahmed:2004,Araki:2005,Aliu:2005}. Depending on the model, Dirac and/or Majorana-type mass terms can be introduced by the couplings of LH and RH neutrinos. With only Dirac-type terms, small neutrino masses can be realized as long as we allow very small Yukawa couplings, $O(10^{-13})$~\cite{Chang:1986bp,Mohapatra:1987hh,Fujikawa:1980yx,Cheng:1980qt}~\footnote{However, with only Dirac-type couplings the active neutrinos are an admixture of left-right neutrinos, making the RH sneutrino NLSP lifetime very short and thus the multicomponent DM scenario not viable. Notice that in the NMSSM+RHN a Majorana mass term is generated dynamically, but in the MSSM+RHN, unless one includes a mass term by hand, neutrino masses are generated only through Dirac couplings.}. On the other hand, considering RH neutrinos with large Majorana masses (up to GUT scale), the neutrino mass pattern can be obtained through a seesaw mechanism with neutrino Yukawa coupling from the order of the electron Yukawa coupling to $O(1)$~\cite{Yanagida:1979as,Ramond:1979py,GellMann:1980vs,Yanagida:1980xy,Glashow1980}. As expected, different realizations imply different LH-RH sneutrino mixing angle ranges.

In the MSSM+RHN, viable thermal RH sneutrino DM can be obtained with large SUSY breaking sneutrino trilinear parameters and significant LH-RH sneutrino mixing~\cite{Arkani:2001,Arina:2007,Thomas:2007bu,Belanger:2010,Belanger:2011ny,Dumont:2012,Kakizaki:2015}. However, if sneutrino trilinear parameters are not large (for example considering that they are suppressed by a small neutrino Yukawa coupling) RH sneutrinos are never in thermal equilibrium in the early Universe~\cite{Gopalakrishna:2006}. Nevertheless, they still could be non-thermal DM (for example produced by the decay of the NLSP after its freeze out), which in turn can result in extremely small LH-RH sneutrino mixing since this parameter depends on the small Yukawa coupling~\cite{Asaka:2006,Gopalakrishna:2006,Asaka:2007,Page:2007,Ishiwata:2009gs,Banerjee:2016,Choi:2018,Ghosh:2019}. Additionally, an alternative to large sneutrino trilinear parameters with thermal RH sneutrino DM can be achieved in models with an extended gauge group~\cite{Lee:2007mt,Bandyopadhyay:2011qm,Belanger:2011rs,Romeri:2012}.

On the other hand, in the NMSSM+RHN the viability of thermal RH sneutrino DM including direct and indirect detection experiments is discussed in Refs.~\cite{Cerdeno:2009,Cerdeno2:2009,Demir:2010,Cerdeno:2011,Cerdeno2:2014,Cerdeno:2015,Cerdeno:2016,Chatterjee:2014,Cao:2019,Lopez-Fogliani:2021qpq,Kim:2021suj}, considering collider and Higgs physics in Refs.~\cite{Wang:2013,Cerdeno:2014,Cerdeno:2018,Cao:2019ofo}, and in the context of spontaneous $R$-parity or CP violation in Refs.~\cite{Kitano:2000,Huitu:2012,Huitu:2014,Tang:2015}. Since the NMSSM includes a singlet superfield to solve dynamically the $\mu$-problem of the MSSM, extra terms with respect to the latter model appear in the superpotential between the RH neutrinos and the singlet superfield. Besides providing a Majorana mass source useful to generate neutrino masses, the phenomenology of the NMSSM+RHN turns out to be very different to the MSSM+RHN. First, new RH sneutrino decay and annihilation channels through the direct coupling to the singlet and singlino appear relating the RH sneutrino sector with CP-even Higgs bosons and neutralino sectors. Therefore, a different DM mass range can be covered and distinctive signatures are predicted for DM and collider experiments. Last but not least, the mass matrix of the sneutrino sector includes new terms involving the scalar singlet vacuum expectation value (VEV) after electroweak symmetry breaking (EWSB). Then, the model presents different LH-RH sneutrino mixing values with respect to the mentioned MSSM extensions.

A comprehensive study and scan of the NMSSM+RHN considering all the latest experimental constraints, with a discussion of the viability, mechanisms and channels used by the RH sneutrino to be a good thermal DM candidate can be found in the recent analyzes Refs.~\cite{Lopez-Fogliani:2021qpq,Kim:2021suj}. In this work, we will consider the solutions from those scans to estimate the LH-RH sneutrino mixing angle range, and to analyze the expected neutrino signal and its prospects of detection in the context of a multicomponent DM scenario.

We organize the paper as follows. In Section~\ref{sec:modelandscan} we present the characteristics of the MSSM+RHN and NMSSM+RHN and focus on the mixing angle in the sneutrino sector. In Section~\ref{sec:sneutgravDM} we discuss the decay channels, relic density and neutrino signal in a coexisting DM scenario with RH sneutrino NLSP and gravitino LSP. In Section~\ref{sec:results} we show the current constraints on neutrino searches with energy $O(\text{GeV})$ by Super-Kamiokande, Antares, and IceCube, and discuss the prospects of detection including next generation of neutrino detectors. The conclusions are left for Section~\ref{sec:conclusions}.

\section{The Models}
\label{sec:modelandscan}

\subsection{MSSM extended with RH neutrino superfields (MSSM+RHN)}
\label{sec:MSSMplusRH}

We will briefly summarize the well-known MSSM including in its formulation RH neutrino superfields to explain the neutrino mass pattern. The aim of this subsection is to describe the most relevant definitions and magnitudes in order to compare them with the NMSSM+RHN.

The MSSM superpotential including a neutrino Dirac- and a Majorana-type mass term is
\bea
W &=& \epsilon_{\alpha \beta} \biggl( Y_e^{ij} \, \hat{H}_d^{\alpha} \, \hat{L}_i^{\beta} \, \hat{e}_j \, + \, Y_d^{ij} \, \hat{H}_d^{\alpha} \, \hat{Q}_i^{\beta} \, \hat{d}_j \, + \, Y_u^{ij} \, \hat{Q}_i^{\alpha} \, \hat{H}_u^{\beta} \, \hat{u}_j \, + \, Y_N^{ij} \, \hat{L}_i^{\alpha} \, \hat{H}_u^{\beta} \, \hat{N}_j \biggr) \nonumber \\ & + & \mu \, \hat{H}_u^{\alpha} \, \hat{H}_d^{\beta} \, + \, \frac{1}{2} \, M_N^{ij} \, \hat{N}_i \, \hat{N}_j,
\label{WMSSMplusRHN}
\eea
where $\epsilon_{\alpha \beta}$ ($\alpha, \beta = 1, 2$) is a totally antisymmetric tensor with $\epsilon_{12} =1$, $i,j = 1,2,3$ are the family indices, and $\hat{N}$ (L=1) the neutrino superfield. The soft SUSY breaking terms are
\bea
V_{soft}&=& \biggl[ \epsilon_{\alpha\beta}  \, \biggl( A_e^{ij} \, Y_e^{ij} \, H_d^{\alpha} \, \tilde{L}_i^{\beta} \, \tilde{e}_j \, + \, A_d^{ij} \, Y_d^{ij} \, H_d^{\alpha} \, \tilde{Q}_i^{\beta} \, \tilde{d}_j \, + \, A_u^{ij} \, Y_u^{ij} \, \tilde{Q}_i^{\alpha} \, H_u^{\beta} \, \tilde{u}_j \, + \, A_N^{ij} \, Y_N^{ij} \, \tilde{L}_i^{\alpha} \, H_u^{\beta} \, \tilde{N}_j \biggr) \nonumber \\ & + & b_{\mu} \, \mu \, H_u^{\alpha} \, H_d^{\beta} \, + \, \frac{1}{2} \, b_{N}^{ij} \, M_N^{ij} \, \tilde{N}_i \, \tilde{N}_j \biggr] \, + h.c. \nonumber \\ & + & m^2_{\phi_{ij}} \, \phi_i^{\dagger} \, \phi_j \, + \, m^2_{\theta_{ij}} \, \theta_i \, \theta_j^* \, + \, m^2_{H_d} \, H_d^{\dagger} \, H_d \, + \, m^2_{H_u} \, H_u^{\dagger} \, H_u \nonumber \\ & + & \frac{1}{2} \, M_1 \, \tilde{B} \, \tilde{B} \, + \, \frac{1}{2} \, M_2 \, \tilde{W}^i \, \tilde{W}^i \, + \, \frac{1}{2} \, M_3 \, \tilde{g}^a \, \tilde{g}^a,
\label{VMSSMplusRHNsoft}
\eea
where $\phi ={\tilde{L},\tilde{Q}}$; $\theta ={\tilde{e},\tilde{N},\tilde{u},\tilde{d}}$ are the scalar components of the corresponding superfields, and the gauginos $\tilde{B}, \tilde{W}, \tilde{g}$, are the fermionic superpartners of the $B$, $W$ bosons, and gluons respectively.

In this work we take all sfermion soft masses and Majorana masses diagonal, $m^2_{ij}=m^2_{ii}=m^2_i$ and vanishing otherwise, without the summation of repeated index convention. Regarding the Yukawa and trilinear couplings, we assume that only the third generation of sfermions are non-zero, $A^{ij}Y^{ij} = T^{ij}$, without the summation convention, except in the neutrino case where $Y_{N}^{ij}$ and $A_N^{ij}$ are taken diagonal.

After the electroweak symmetry breaking (EWSB) induced by the soft SUSY-breaking terms of $O(\text{TeV})$, and with the choice of CP conservation, the neutral Higgses ($H_{u,d}$) develop the following vacuum expectation values (VEVs)
\begin{equation}
\langle H_d \rangle = \frac{v_d}{\sqrt{2}}, \hspace{1cm} \langle H_u \rangle = \frac{v_u}{\sqrt{2}},
\label{vevs1}
\end{equation}
where $v^2=v_d^2+v_u^2= 4m_Z^2 /(g^2 + g'^2) \simeq (246 \text{ GeV})^2$, with $m_Z$ the $Z$ boson mass, and $g$ and $g'$ the $U(1)_Y$ and $SU(2)_L$ coupling respectively.

\subsection{NMSSM extended with RH neutrino superfields (NMSSM+RHN)}
\label{sec:NMSSMplusRH}

The NMSSM+RHN can explain the neutrino mass pattern, and solves the $\mu$-problem of the MSSM and MSSM+RHN. The NMSSM+RHN superpotential is
\bea
W &=& \epsilon_{\alpha \beta} \biggl( Y_e^{ij} \, \hat{H}_d^{\alpha} \, \hat{L}_i^{\beta} \, \hat{e}_j \, + \, Y_d^{ij} \, \hat{H}_d^{\alpha} \, \hat{Q}_i^{\beta} \, \hat{d}_j \, + \, Y_u^{ij} \, \hat{Q}_i^{\alpha} \, \hat{H}_u^{\beta} \, \hat{u}_j \, + \, Y_N^{ij} \, \hat{L}_i^{\alpha} \, \hat{H}_u^{\beta} \, \hat{N}_j \nonumber \\ & + & \lambda \, \hat{S} \, \hat{H}_u^{\alpha} \, \hat{H}_d^{\beta} \biggr) \, + \, \lambda_N^{ij} \, \hat{S} \, \hat{N}_i \, \hat{N}_j \, + \, \frac{\kappa}{3} \, \hat{S} \, \hat{S} \, \hat{S},
\label{WNMSSMplusRHN}
\eea
where $\hat{S}$ (L=0) is a singlet superfield, and $\hat{N}$ (L=1) the neutrino superfield. In this case, a $Z_3$ symmetry is introduced to forbid the appearance of any dimensional parameters. The soft SUSY breaking terms are
\bea
V_{soft}&=& \biggl[ \epsilon_{\alpha\beta}  \, \biggl( A_e^{ij} \, Y_e^{ij} \, H_d^{\alpha} \, \tilde{L}_i^{\beta} \, \tilde{e}_j \, + \, A_d^{ij} \, Y_d^{ij} \, H_d^{\alpha} \, \tilde{Q}_i^{\beta} \, \tilde{d}_j \, + \, A_u^{ij} \, Y_u^{ij} \, \tilde{Q}_i^{\alpha} \, H_u^{\beta} \, \tilde{u}_j \, + \, A_N^{ij} \, Y_N^{ij} \, \tilde{L}_i^{\alpha} \, H_u^{\beta} \, \tilde{N}_j \nonumber \\ & + & A_{\lambda} \, \lambda \, S \, H_u^{\alpha} \, H_d^{\beta} \biggr) \, + \, A_{\lambda_N}^{ij} \, \lambda_N^{ij}  \, S \, \tilde{N}_i \, \tilde{N}_j \, + \, \frac{A_{\kappa} \, \kappa}{3} \, S \, S \, S \biggr] \, + h.c. \nonumber \\ & + & m^2_{\phi_{ij}} \, \phi_i^{\dagger} \, \phi_j \, + \, m^2_{\theta_{ij}} \, \theta_i \, \theta_j^* \, + \, m^2_{H_d} \, H_d^{\dagger} \, H_d \, + \, m^2_{H_u} \, H_u^{\dagger} \, H_u \, + \, m^2_S \, S \, S^* \nonumber \\ & + & \frac{1}{2} \, M_1 \, \tilde{B} \, \tilde{B} \, + \, \frac{1}{2} \, M_2 \, \tilde{W}^i \, \tilde{W}^i \, + \, \frac{1}{2} \, M_3 \, \tilde{g}^a \, \tilde{g}^a,
\label{VNMSSMplusRHNsoft}
\eea
where $S$ is the scalar component of the single superfield $\hat{S}$. As in the MSSM extension, we consider the sfermion soft masses diagonal, as well as the Yukawa and trilinear couplings. Furthermore, we also consider diagonal parameters $\lambda_N^{ij}=\lambda_N^{ii}=\lambda_N^{i}$, and its corresponding trilinear coupling, $A_{\lambda_N}^{ij} \, \lambda_N^{ij}=A_{\lambda_N}^{i} \, \lambda_N^{i} = T_{\lambda_N}^{i}$.

After the electroweak symmetry breaking (EWSB) induced by the soft SUSY-breaking terms of $O(\text{TeV})$, in addition to Eq.~(\ref{vevs1}) the scalar singlet develops a VEV
\begin{equation}
\langle s \rangle = \frac{v_s}{\sqrt{2}}.
\label{vevs2}
\end{equation}

To generate an effective $\mu$-term, $\mu_{eff}=\frac{\lambda \, v_s}{\sqrt{2}}$, $v_s=O(\text{GeV-TeV})$ is needed. Then it follows that the superpotential term $\lambda_N^{i}  \, S \, \hat{N}_i \, \hat{N}_i$ generates dinamically a RH neutrino Majorana mass term $M_{N}^{i}=\frac{\lambda_N^{i} \, v_s}{\sqrt{2}} \sim O(\text{GeV-TeV})$, assuming that the parameters $\lambda$ and $\lambda_N^{i}$ are $O(0.1 - 1)$.

The superpotential of the NMSSM+RHN  in Eq.~(\ref{WNMSSMplusRHN}) as well as in the NMSSM case has a $Z_3$ symmetry. It is well known that this discrete symmetry when spontaneously broken can induce a cosmological domain wall problem~\cite{Holdom:1983vk,Ellis:1986mq,Rai:1992xw,Abel:1995wk}. However, this problem can be solved via the presence of non-renormalizable operators in the superpotential $W$. These operators break explicitly the $Z_3$ symmetry lifting the degeneracy of the three original vacua, and they can be chosen small enough not to alter the low-energy phenomenology~\cite{Holdom:1983vk,Ellis:1986mq,Rai:1992xw}. In the context of supergravity they can reintroduce in $W$ the linear and bilinear terms forbidden by the $Z_3$ symmetry~\cite{Abel:1995wk}, and generate quadratic tadpole divergences~\cite{Ellwanger:1983mg,Bagger:1993ji,Jain:1994tk,Bagger:1995ay,Abel:1996cr}, nevertheless these problems can be eliminated in models with a $R$-symmetry in the non-renormalizable superpotential~\cite{Abel:1996cr,Panagiotakopoulos:1998yw}.

\subsection{Neutrino sector}
\label{sec:neutrinosector}

In the MSSM+RHN case, if we do not introduce a mass scale in the superpotential and consider pure Dirac-type neutrinos, to reproduce the measured LH neutrino masses, $m_{\nu_{L}}$, we need~\cite{Asaka:2006,Asaka:2007} 
\bea
Y_N \, \sin \beta \sim 10^{-13} \, \biggl( \frac{m^2_{\nu_{L}}}{2.8\times10^{-3} \text{eV}^2}  \biggr),
\label{yukawaMSSM}
\eea
with $m_D \simeq Y_N \, v_u = Y_N \, v \, \sin \beta$ the Dirac mass and $\tan \beta = \langle H_u \rangle / \langle H_d \rangle$. Assuming a relatively large $\tan \beta$ such that $H_u$ behaves like the SM Higgs, we require $Y_N \sim 10^{-13}$.

As we will see in Section~\ref{sec:31}, for RH sneutrino NLSP and gravitino LSP, the lifetime of the former would be much shorter than the age of the Universe (unless we fine-tune its masses), if the RH sneutrino decay channel to gravitino plus RH neutrino is allowed. For pure Dirac-type neutrinos this channel is always open, therefore a coexisting DM scenario would not be possible.\\

If the LH neutrino masses are generated by a seesaw mechanism the general neutrino mass matrix is given by
\begin{equation}
M_{\nu} = 
\left(\begin{array}{cc}
 0 & m_D\\
  m_D^T & M_N
  \end{array}\right),
\end{equation}
so,
\bea
m_{\nu_{L}}\simeq - m_D \, M_N^{-1} \, m_D^T, 
\eea

In the MSSM+RHN case a Type-I seesaw results for different choices of $M_N$ and $Y_N$. For example, $M_N\sim 10^{16}$ GeV with $Y_N \sim O(1)$, or $M_N\sim O(\text{EW})$ with $Y_N \sim 10^{-6}$. Additionally, the inclusion of a Majorana term results in heavy RH neutrino masses, $m_{\nu_{R}}$. Then, the problematic decay channel mentioned above can be kinematically forbidden and the multicomponent DM scenario is possible if $m_{\nu_{R}} > m_{\tilde{\nu}_{R}}$, with $m_{\tilde{\nu}_{R}}$ the RH sneutrino mass.

However, assuming weak-scale trilinear sneutrino couplings proportional to the Yukawa, $T_N=A_N Y_N$ with $A_N \sim O(\text{EW})$, in Refs.~\cite{Gopalakrishna:2006,Ghosh:2019} it has been shown that to avoid RH sneutrino DM overabundance, the Majorana scale should be much smaller than a few tens of GeV. Within that framework, one would get $m_{\nu_{R}} \lesssim O(10\text{GeV})$, which is lighter than the typical allowed RH sneutrino mass range found in MSSM+RHN analysis, $m_{\tilde{\nu}_{R}}\gtrsim O(\text{100GeV})$, and again the RH sneutrino would have a short lifetime. On the other hand, if we consider large trilinear sneutrino couplings $T_N \sim O(\text{EW})$ (not suppressed by the Yukawas) thermal DM production is allowed through large LH-RH mixing.

Interestingly, in the NMSSM+RHN thermal RH sneutrino DM in a wide mass range is viable due to its couplings with the singlet field. Furthermore, a Majorana mass term does not need to be introduced by hand since an effective term is dynamically generated at the EW scale, $M_{N}^{i}=\frac{\lambda_N^{i} \, v_s}{\sqrt{2}} \sim O(\text{GeV-TeV})$, in order to generate an effective $\mu$-term. Thus with the seesaw mechanism
\bea
m_{\nu_{L}} \simeq \frac{Y_N^2 \, v_u^{2}}{\lambda_N \, v_s} \sim Y_N^2 \; \times \text{EW scale}, 
\label{neutrinomassNMSSMplusRHN}
\eea
which implies $Y_N \sim 10^{-6}$ to get the right order for the neutrino masses.

\subsection{Sneutrino sector}
\label{sec:sneutrinosector}

The sneutrinos form a $12\times 12$ mass matrix divided into $6 \times 6$ submatrices
\begin{equation}
M_{\tilde{\nu}}^2 = 
\left(\begin{array}{cc}
 m_{\mathbb{R}\mathbb{R}}^2 & 0_{6\times 6}\\
  0_{6\times 6} & m_{\mathbb{I}\mathbb{I}}^2
  \end{array}\right),
\end{equation}
where the subscripts $\mathbb{R}$ and $\mathbb{I}$ denote CP-even and CP-odd states respectively. The off-diagonal submatrices are zero due to our choise of CP conservation. The submatrices are
\begin{equation}
M_{\mathbb{R}\mathbb{R}}^2 = 
\left(\begin{array}{cc}
 m_{L_{i}}^2 & A_{i}^{+}\\
  (A_{i}^{+})^T & m_{R_{i}}^2 + B_i
  \end{array}\right),
\end{equation}
\begin{equation}
M_{\mathbb{I}\mathbb{I}}^2 = 
\left(\begin{array}{cc}
 m_{L_{i}}^2 & A_{i}^{-}\\
  (A_{i}^{-})^T & m_{R_{i}}^2 - B_i
  \end{array}\right),
\end{equation}
with
\bea
& & \text{MSSM+RHN} \hspace{4cm} \text{NMSSM+RHN} \nonumber \\
A_i^{+} &=& Y_N^{i} \, \left( A_N^i \, v_u \, + \, M_N^i \, v_u \,  - \, \mu \, v_d \right), \hspace{1cm} Y_N^{i} \, \left( A_N^i \, v_u \, + \, 2 \, \lambda_N^{i} \, v_u \, v_s \, - \, \lambda \, v_d \, v_s \right),\label{aplus}\\
A_i^{-} &=& Y_N^{i} \, \left( A_N^i \, v_u \, - \, M_N^i \, v_u \,  - \, \mu \, v_d \right), \hspace{1cm} Y_N^{i} \, \left( A_N^i \, v_u \, - \, 2 \, \lambda_N^{i} \, v_u \, v_s \, - \, \lambda \, v_d \, v_s \right),\label{aminus}\\
B_i &=& -b_N^{i} \, M_N^{i}, \hspace{4.8cm} 2 \, \lambda_N^{i} \, \left( \, A_{\lambda_N}^{i} \, v_s \, + \, \kappa \, v_s^{2} \, - \, \lambda  \, v_u \, v_d \, \right),\label{bsneutrino}\\
m_{L_i}^2 &=& m_{\tilde{L}_i}^2 \, + \, (Y_N^{i})^2 \, v_u^2 \, + \, \frac{1}{2} \, m_Z^2 \, \cos 2 \beta, \hspace{0.7cm} m_{\tilde{L}_i}^2 \, + \, (Y_N^{i})^2 \, v_u^2 \, + \, \frac{1}{2} \, m_Z^2 \, \cos 2 \beta,\\
m_{R_i}^2 &=& m_{\tilde{N}_i}^2 \, + \, (Y_N^{i})^2 \, v_u^2 \, + \, (M_N^{i})^2, \hspace{1.7cm} m_{\tilde{N}_i}^2 \, + \, (Y_N^{i})^2 \, v_u^2 \, + \, 4 \, (\lambda_N^{i})^2 \, v_s^2,
\eea
as before, $i = 1,2,3$ is the family index.

From Eqs.~(\ref{aplus}) and (\ref{aminus}) we can see that the LH-RH mixing is proportional to the Yukawa coupling $Y_N$. If we assume a small LH-RH mixing, the LH and RH sneutrino eigenstates correspond to the respective mass eigenstates
\bea
m_{\tilde{\nu}_{L_i}}^2 \simeq m_{\tilde{L}_i}^2 \, + \, \frac{1}{2} \, m_Z^2 \, \cos 2 \beta , \hspace{1cm} m_{\tilde{\nu}_{R_i}}^2 \simeq m_{R_i}^2 \, \pm \, B_i,
\label{aproxLHRHmasses}
\eea
here also the upper (lower) sign corresponds to the CP-even (CP-odd) state. Eq.~(\ref{bsneutrino}) shows that the RH sneutrino mass splitting is proportional to the Majorana mass parameters. For typical parameter values
\bea
\begin{array}{cc}
\text{MSSM+RHN} \hspace{1cm} & m_{\tilde{\nu}_{R_i}}^2 \sim m_{\tilde{N}_i}^2 + (M_N^i)^2 \mp (b_N^i M_N^i), \\
\text{NMSSM+RHN} \hspace{1cm} & m_{\tilde{\nu}_{R_i}}^2 \sim m_{\tilde{N}_i}^2 + (2 \, \lambda_N^{i} \, v_s)^2 \pm \lambda_N^{i} \, \left( 2 \, A_{\lambda_N}^{i} \, v_s \, + \, 2 \, \kappa \, v_s^{2} \right).
\end{array}
\label{RHsneutrinomasses}
\eea
Notice that in the NMSSM+RHN, the squared RH sneutrino mass is proportional to $\lambda_N^i$, therefore if $\lambda_N \rightarrow 0$ then $m_{\tilde{\nu}_{R_i}}^2 \simeq  m_{\tilde{N}_i}^2$.

\subsection{LH and RH mixing angle}
\label{sec:LHRHmixing}

Considering only one family of neutrinos, the mixing angle between LH and RH sneutrinos, $\theta_{\tilde{\nu}}$, can be approximated by
\bea
\tan 2 \theta_{\tilde{\nu}} & \simeq & \frac{2 \, A^{\pm}}{m_{L}^2 \, - \, \left( m_{R}^2 \, \pm \, B \right)}, \label{sneutmix1}
\eea
where the upper (lower) sign corresponds to the CP-even (CP-odd) state.\\


To get thermal RH sneutrino DM in the MSSM+RHN one can consider weak-scale trilinear sneutrino couplings, $T_N=A_N Y_N \sim O(\text{EW})$, i.e. not suppressed by the small neutrino Yukawa coupling, inducing large LH-RH mixing (see Eqs.~(\ref{aplus}) and (\ref{aminus}))~\cite{Arkani:2001,Arina:2007,Thomas:2007bu,Belanger:2010,Belanger:2011ny,Dumont:2012,Kakizaki:2015}
\bea
\tan 2 \theta_{\tilde{\nu}}  \simeq \frac{2 \, T_N \, v_u }{ m_{\tilde{\nu}_{L}}^2 \, - \,  m_{\tilde{\nu}_{R}}^2 } \sim O(0.01-1), 
\label{sneutrinomixMSSMTH}
\eea
where in the last term we consider typical EW scale parameters. In this way, the RH sneutrino LSP is not sterile but couples to the SM gauge and Higgs bosons through the mixing with its LH counterpart, and therefore is a viable thermal DM candidate.\\

In the NMSSM+RHN case the mixing can be expressed as
\bea
\tan 2 \theta_{\tilde{\nu}} \simeq \frac{2 \, Y_N \, \left( \, A_N \, v_u \, \pm \, 2 \, \lambda_N \, v_u \, v_s \, - \, \lambda \, v_d \, v_s \, \right)}{m_{\tilde{L}}^2 \, + \, \frac{1}{2} \, m_Z^2 \, \cos 2 \beta \, - \, m_{\tilde{N}}^2 \, - \, 4 \, \lambda_N^2 \, v_s^2 \, \mp \, 2 \, \lambda_N \, \left( \, A_{\lambda_N} \, v_s \, + \, \kappa \, v_s^{2} \, - \, \lambda  \, v_u \, v_d \, \right)}. \label{sneutrinomixNMSSM}
\eea

Lets consider some special cases. Taking $\lambda_N \rightarrow 0$ we recover the MSSM+RHN result. If we allow large SUSY breaking sneutrino trilinear parameters $Y_N \, A_N \sim O(\text{GeV})$, we get $\tan 2 \theta_{\tilde{\nu}} \sim O(0.1)$.

With a non-vanishing $\lambda_N$ and for typical parameter values, $A_N \sim A_{\lambda_N} \sim O(\text{GeV})$, $\tan \beta \sim O(10)$, $\lambda \sim \kappa \sim \lambda_N$, we get
\bea
\tan 2 \theta_{\tilde{\nu}} & \sim & \frac{ Y_N \, \lambda_N \, v_s \, v_u}{m_{\tilde{L}}^2 \, + \, \frac{1}{2} \, m_Z^2 \, \cos 2 \beta \, - \, m_{\tilde{N}}^2 \, - \, \lambda_N^2 \, v_s^2 } \sim 10^{-2} \times Y_N  \sim O(10^{-8}), \label{sneutmix2}
\eea
where we have used that $m_{\tilde{L}} \sim O(10^2 - 10^3)$ GeV, and that $\lambda_N \, v_s \sim O(\text{EW})$ with $Y_N\sim 10^{-6}$ to reproduce the neutrino masses. Remarkably, in the NMSSM+RHN the natural sneutrino mixing angle value (without taking an extremely small neutrino Yukawa nor too large trilinear neutrino couplings) differs by several orders of magnitude with respect to the MSSM+RHN case, satisfying the neutrino physics and having the lightest RH sneutrino as a viable thermal DM candidate.

In order to compute numerically the range of the LH and RH sneutrino mixing angle we scan the low-energy parameter space of the NMSSM+RHN as described in the recent Refs.~\cite{Lopez-Fogliani:2021qpq,Kim:2021suj}. We used the \texttt{MultiNest}~\cite{multinestcite} algorithm as optimizer, the package \texttt{SARAH}~\cite{sarahcite1,sarahcite2,sarahcite3} to build the model, and \texttt{SPheno}~\cite{sphenocite1,sphenocite2} to generate the particle spectrum and other observables such as flavour and $(g-2)_{\mu}$. \texttt{MicrOmegas}~\cite{micromegascite1,micromegascite2, micromegascite3} is used to compute the relic density of the RH sneutrino NLSP since in our scenario the gravitino LSP is decoupled from the rest of the spectrum and therefore does not affect the evolution of the other thermal relics. \texttt{HiggsBounds}~\cite{HBcite1,HBcite2,HBcite3} and \texttt{HiggsSignals}~\cite{HScite1,HScite2,HScite3} are used to determine the compatibility of the Higgs sector against current constraints and measurements, \texttt{DDCalc}~\cite{ddcalccite} takes into account direct detection bounds, and  \texttt{SModelS}~\cite{Kraml:2013mwa} is employed to impose 8 TeV and 13 TeV new physics searches performed by the ATLAS and CMS experiments (see Appendix~\ref{appen0} 
for further details on scan methodology and constraint considered).

Notice that in the numerical computations done for this work, we assume a diagonal Yukawa matrix with $Y_N^{ii} = 10^{-6}$, with $i=1,...,3$, which can not reproduce the measured values of the neutrino sector (mixing angles and mass differences). To satisfy those constraints, for every solution an extra scan would be needed fixing every parameter but $Y_N^{ij}$ with $i,j=1,...,3$. This is possible because the DM and neutrino sectors are decoupled and one of our goals is to estimate the LH-RH mixing angle range (see Appendix~\ref{appendixneutrino}).

\begin{figure}[t!]
\begin{center}
 \begin{tabular}{cc}
 \hspace*{-14mm}
 \epsfig{file=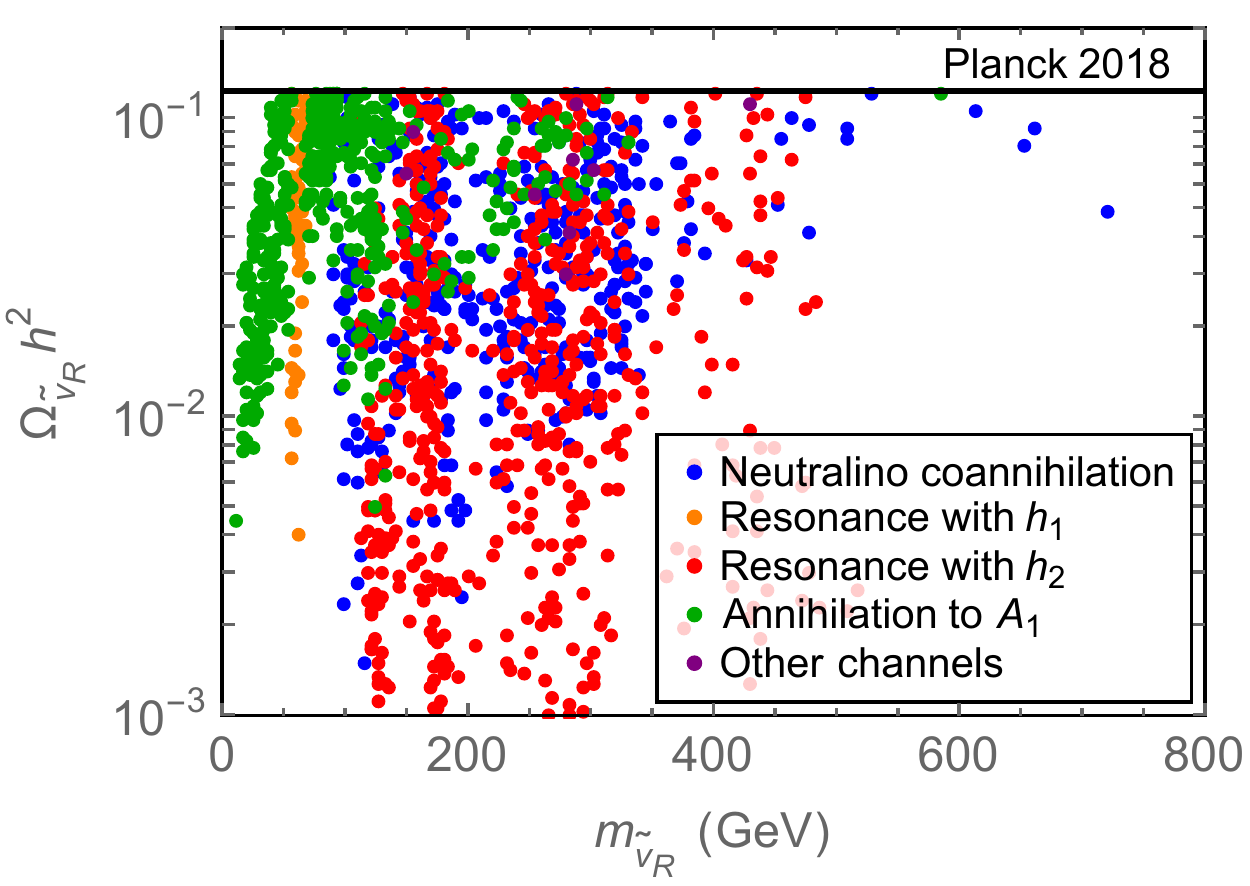,height=6.5cm} 
 \hspace*{-5mm} \epsfig{file=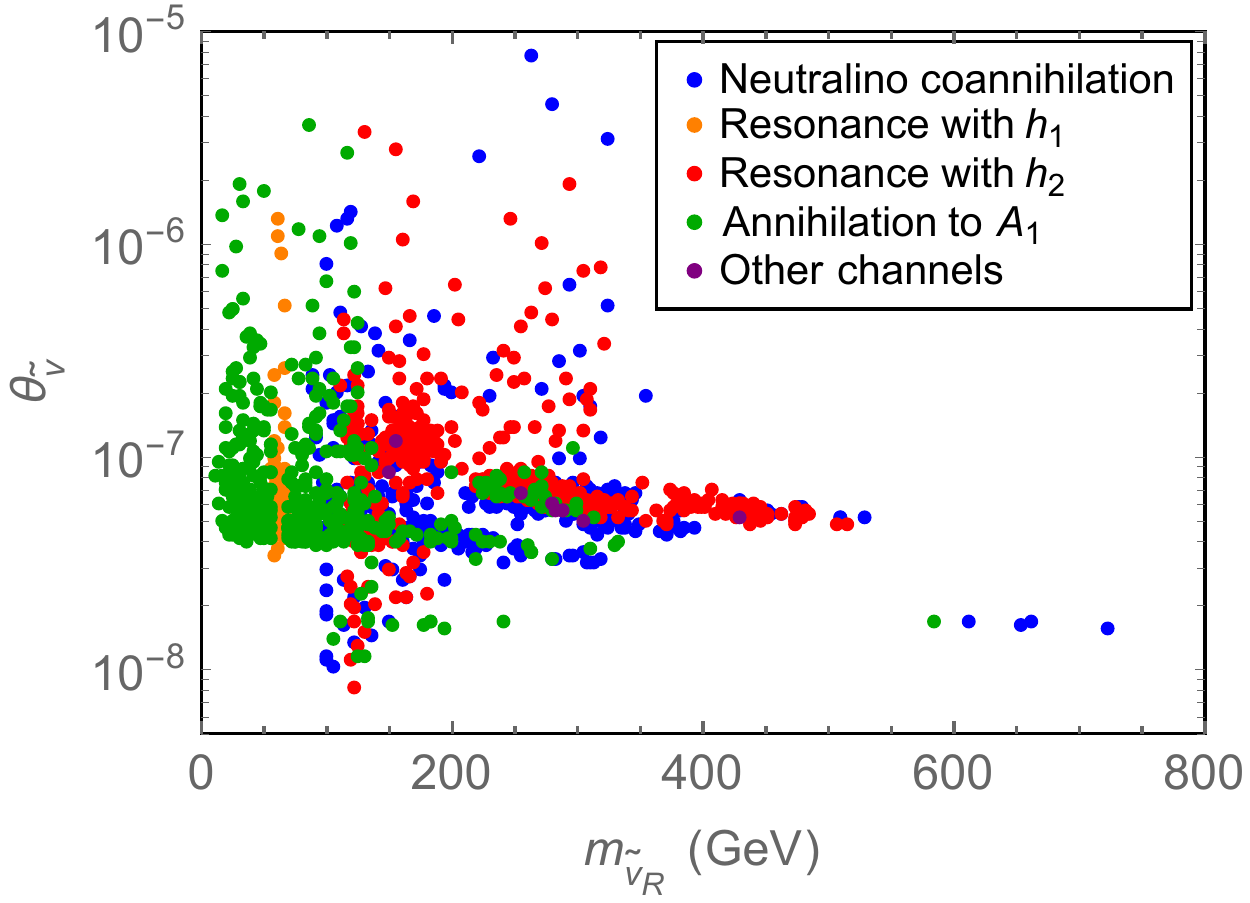,height=6.5cm} 
    \end{tabular}
    \captions{Relic density and mixing angle in the sneutrino sector versus RH sneutrino mass. The color coding represents the main channels used by RH sneutrinos to obtain an allowed relic density, with $h_i$ ($A_i$) a neutral CP-even (CP-odd) scalar of the Higgs-singlet sector.
}
    \label{sneutgrav-scan}
\end{center}
\end{figure}

In Fig.~\ref{sneutgrav-scan} we show the points that satisfy all the constraints implemented considering solutions with RH sneutrino as DM candidate with a significant contribution to the total DM relic density (greater than 1\% of the current measured value). Notice that the mixing angle in the sneutrino sector lies within the expected range:
\begin{equation}
10^{-8} \leq \theta_{\tilde{\nu}} \leq 10^{-6},
\label{sintheta_range1}
\end{equation}
for $10 \text{ GeV} \lesssim m_{\tilde{\nu}_R} \lesssim 800 \text{ GeV}$.

Finally, we would like to highlight that solutions with low mass RH sneutrino LSP are possible in the NMSSM+RHN due to the presence of very light pseudo-scalars with dominant singlet composition. The main channel to obtain a consistent relic density in this region involves the term $\lambda_N S N N$, absent in the MSSM extended with RH neutrino superfields.

\section{RH sneutrino and gravitino Dark Matter}
\label{sec:sneutgravDM}

In this section we discuss the decay channels, relic density and the characteristic neutrino signal in a coexisting DM scenario with RH sneutrino NLSP and gravitino LSP. We present
current constraints
on its masses and the reheating temperature of the Universe considering neutrino searches by Super-Kamiokande, Antares, and IceCube. We also include prospect of detection by current and next generation of neutrino detectors.

\subsection{RH sneutrino NLSP decay to gravitino LSP and relic density}
\label{sec:31}

In the framework of supergravity, the gravitino has an R-parity conserving interaction term with the RH sneutrino and the RH neutrino. If the gravitino is the LSP and the RH sneutrino the NLSP, the NLSP decay is forbidden kinematically since RH neutrinos are usually heavier. However, it is possible with active neutrinos in the final state through mixing between LH and RH sneutrinos. Then, the decay width becomes~\cite{Feng:2004}
\bea
\Gamma(\tilde{\nu}_R \rightarrow \Psi_{3/2} \, \nu_L) = \frac{1}{48 \, \pi \, M_P^2} \,  \frac{m^5_{\tilde{\nu}_R}}{m_{3/2}^2} \, \left( 1 - \frac{m_{3/2}^2}{m_{\tilde{\nu}_R}^2} \right)^4
 \, \sin^2 \theta_{\tilde{\nu}},
 \label{sneutgravdecaywidth}
\eea
where $\tilde{\nu}_R$, $\Psi_{3/2}$ and $\nu_L$ denote the RH sneutrino, the gravitino and the active neutrinos respectively, $m_{\tilde{\nu}_R}$ the RH sneutrino mass, $m_{3/2}$ the gravitino mass, and $M_P=2.4 \times 10^{18}$ GeV the reduced Planck mass scale.

As we can see in Eq.~(\ref{sneutgravdecaywidth}), the RH sneutrino decay is suppressed both by the small LH-RH mixing angle $\theta_{\tilde{\nu}}$ and by the scale of the gravitational interaction. Assuming that $m_{\tilde{\nu}_R}\gg m_{3/2}$, the lifetime of the NLSP can be approximated by
\bea
\tau_{\tilde{\nu}_R}\simeq \Gamma^{-1}(\tilde{\nu}_R \rightarrow \Psi_{3/2} \, \nu_L) \simeq 5.7\times 10^{23} s \, \left( \frac{10 \text{ GeV}}{m_{\tilde{\nu}_R}} \right)^5 \, \left( \frac{m_{3/2}}{0.1 \text{ GeV}} \right)^2 \, \left( \frac{10^{-8}}{\sin \theta_{\tilde{\nu}}} \right)^2.
\label{sneutgravlifetime}
\eea
For RH sneutrino masses $\sim O(\text{GeV})$, the NLSP lifetime is longer than the age of the Universe ($\sim 10^{17}$s), and the energy ranges of neutrino telescopes like Super-Kamiokande, IceCube and Antares make them ideal instruments to measure or set constraints on the neutrino flux produced by its decay.

Notice that if the channel $\tilde{\nu}_R \rightarrow \Psi_{3/2} \, N_i$ is allowed, with $N_i$ a RH neutrino, the RH sneutrino lifetime would be typically much shorter than the age of the Universe (unless $m_{\tilde{\nu}_R} \simeq m_{3/2}$) and therefore not a dark matter candidate. However, we have checked that all the NMSSM+RHN solutions shown in this work present a RH neutrino heavier than the RH sneutrino NLSP, hence the problematic channel is kinematically forbidden.

For the two component DM scenario, the total DM relic density would be
\bea
\Omega_{\tilde{\nu}_R}h^2+\Omega_{3/2}h^2=\Omega_{cdm}^{\text{Planck}}h^2,
\label{relicsum3}
\eea
with $\Omega_{cdm}^{\text{Planck}}h^2\simeq 0.1198 \pm 0.0012$ the relic density value observed by the Planck collaboration at recombination~\cite{Planck:2018vyg}, $\Omega_{\tilde{\nu}_R}h^2$ the relic density contribution of RH sneutrinos calculated with \texttt{MicrOmegas}, and $\Omega_{3/2}h^2$ the gravitino relic, whose thermal abundance is expected to be~\cite{Bolz:2000fu,Rychkov:2007uq,Eberl:2020fml}
\begin{equation}
\Omega^{TP}_{3/2}h^2\simeq 0.02\left(\frac{T_R}{10^5 \text{ GeV}}\right)\left(\frac{1 \text{ GeV}}{m_{3/2}}\right)\left(\frac{M_3(T_R)}{3\text{ TeV}}\right)^2\left(\frac{\gamma (T_R) / (T_R^6/M_P^2)}{0.4}\right),
\label{relicgravitinos}
\end{equation}
where $M_3(T_R)$ is the running gluino mass, and the last factor parametrizes the effective production rate raging $\gamma (T_R) / (T_R^6/M_P^2)\simeq 0.4-0.35$, for $T_R\simeq 10^4-10^6$ GeV~\cite{Rychkov:2007uq,Eberl:2020fml}. For our numerical calculations we approximate $\gamma (T_R) / (T_R^6/M_P^2)=0.4$ and $M_3=3$ TeV.

\subsection{Neutrino signal from NLSP to LSP decay}
\label{sec:neutrinosignal}

For the galactic halo, the differential flux of neutrinos from DM decay is calculated by integrating its distribution around us along the line of sight (notice that we can reconstruct the direction of their origin because neutrinos, like photons, are unaffected by magnetic fields)~\cite{Grefe:2008zz}: 
\begin{equation}
 \frac{d\Phi_{\nu}^{\text{halo}}}{dEd\Omega}= \frac{r_{\textit{DM}}}{4\,\pi\,\tau_{\textit{DM}}\,m_{\textit{DM}}}\,\frac{1}{\Delta\Omega}\,\frac{dN^{\text{total}}_{\nu}}{dE} \int_{\Delta\Omega}\!\!\cos
 b\,db\,d\ell\int_0^{\infty}\!\! ds\,\rho_{\text{halo}}(r(s,\,b,\,\ell))\ ,
 \label{eq:decayFlux}
\end{equation}
where $r_{\textit{DM}}$, $\tau_{\textit{DM}}$, $m_{\textit{DM}}$ are the relic density fraction, lifetime, mass of the decaying DM particle respectively, $\frac{dN^{\text{total}}_{\nu}}{dE}$ is the total number of neutrinos produced in a DM decay, $\Delta \Omega$ is the solid angle supported by the region of interest (ROI), i.e. the observed region of the sky, $b$ and $\ell$ denote the Galactic latitude and longitude, respectively, and $s$ denotes the distance from the Solar System. The radius $r$ in the DM halo density profile of the Milky Way, $\rho_{\text{halo}}$, is expressed in terms of these Galactic coordinates.

It is important to point out that we have to recast the published constraints taking into account three key assumptions that the experimental collaborations usually consider:
\begin{itemize}
    \item DM annihilation channels instead of decay channels,
    \item only one DM component instead of multiple candidates,
    \item a pair of massless neutrinos in the final state (or a pair of other SM particles, like $b\bar{b}$ subsequently decaying to neutrinos and other particles) instead of a neutrino and a massive gravitino in the final state.
\end{itemize}

With respect to the first point, in Appendix~\ref{appen1} we discuss in detail and show how to estimate the constraints for decaying channels from the annihilation limits presented by the collaborations.

Once we have the decay channel constraints as upper limits on the DM lifetime, we can focus on the other points. In Eq.~(\ref{eq:decayFlux}) we allow a multicomponent DM scenario and assume that the distribution of each DM component is homogeneous along the DM distribution, i.e. every component `$i$' proportionally follows the same DM density profile, $\rho_{\text{total}}(x) = \sum_i \alpha_i \, \rho_{\text{halo}}(x)$.
The factor $r_{\textit{DM}} =\frac{\Omega_{\text{DM}}}{\Omega_{cdm}^{\text{Planck}}}$ represents that only a fraction of the total DM budget, with relic density $\Omega_{\text{DM}}$, produces neutrinos.

As mentioned, the experimental collaborations usually consider only one DM component, and therefore take $r_{DM}=1$ for their analysis to place constraints on the DM lifetime and mass. In a multicomponent scenario instead of working with a lower flux, it is useful to assume an effective lifetime 
\begin{equation}
\tau_{\text{DM-eff}}=\frac{\tau_{\text{DM}}}{r_{\text{DM}}},
\label{newtime1}
\end{equation}
where $\tau_{\text{DM}}=\Gamma^{-1}(\text{DM}\rightarrow\nu \, ...)$. Then, the effective lifetime $\tau_{\text{DM-eff}}$ could be compared with the reported constraints.

In this work, to describe the neutrino flux we can use Eq.~(\ref{eq:decayFlux}) with the RH sneutrino NLSP taking the role of the decaying DM to gravitino plus neutrinos as long as its lifetime is greater than the age of the Universe (see Eq.~(\ref{sneutgravlifetime})). The RH sneutrino effective lifetime is
\begin{equation}
\tau_{\tilde{\nu}_R\text{-eff}}=\frac{\Gamma^{-1}(\tilde{\nu}_R \rightarrow \Psi_{3/2} \, \nu_L)}{r_{\tilde{\nu}_R}} \hspace{1cm} \text{with} \hspace{1cm} r_{\tilde{\nu}_R}=\frac{\Omega_{\tilde{\nu}_R}}{\Omega_{cdm}^{\text{Planck}}}.
\label{newtime2}
\end{equation}

Finally, to probe our scenario with a neutrino and a gravitino in the final state, we are going to employ the constraints with a pair of neutrinos in the final state, since the gravitino is completely stable and we can consider it decoupled from the rest of the spectrum being its interactions suppressed by the Planck scale. However we have to take into account that the energy of the signal, a monochromatic neutrino, is going to be modified for a massive gravitino. In our scenario the energy of the produced monochromatic neutrino is given by
\bea
E_{\nu}=\frac{m_{\tilde{\nu}_R}^2-m_{3/2}^2}{2m_{\tilde{\nu}_R}}.
\label{neutrinoenergy}
\eea
In the region where the gravitino can be considered massless (i.e. $m_{3/2} \ll m_{\tilde{\nu}_R}$), the energy of the signal would be the same as in the case considered by the experimental collaborations. However, if $m_{3/2}\sim m_{\tilde{\nu}_R}$ then $E_{\nu}\simeq m_{\tilde{\nu}_R}-m_{3/2}$. This means that the published constraints will be shifted since they are usually presented as a function of the DM mass, and not the energy of the signal.

\subsection{Constraints on NMSSM+RHN from neutrino telescopes}
\label{sec:results}

\begin{figure}[t!]
\begin{center}
 \begin{tabular}{c}
 \hspace*{-4mm}
 \epsfig{file=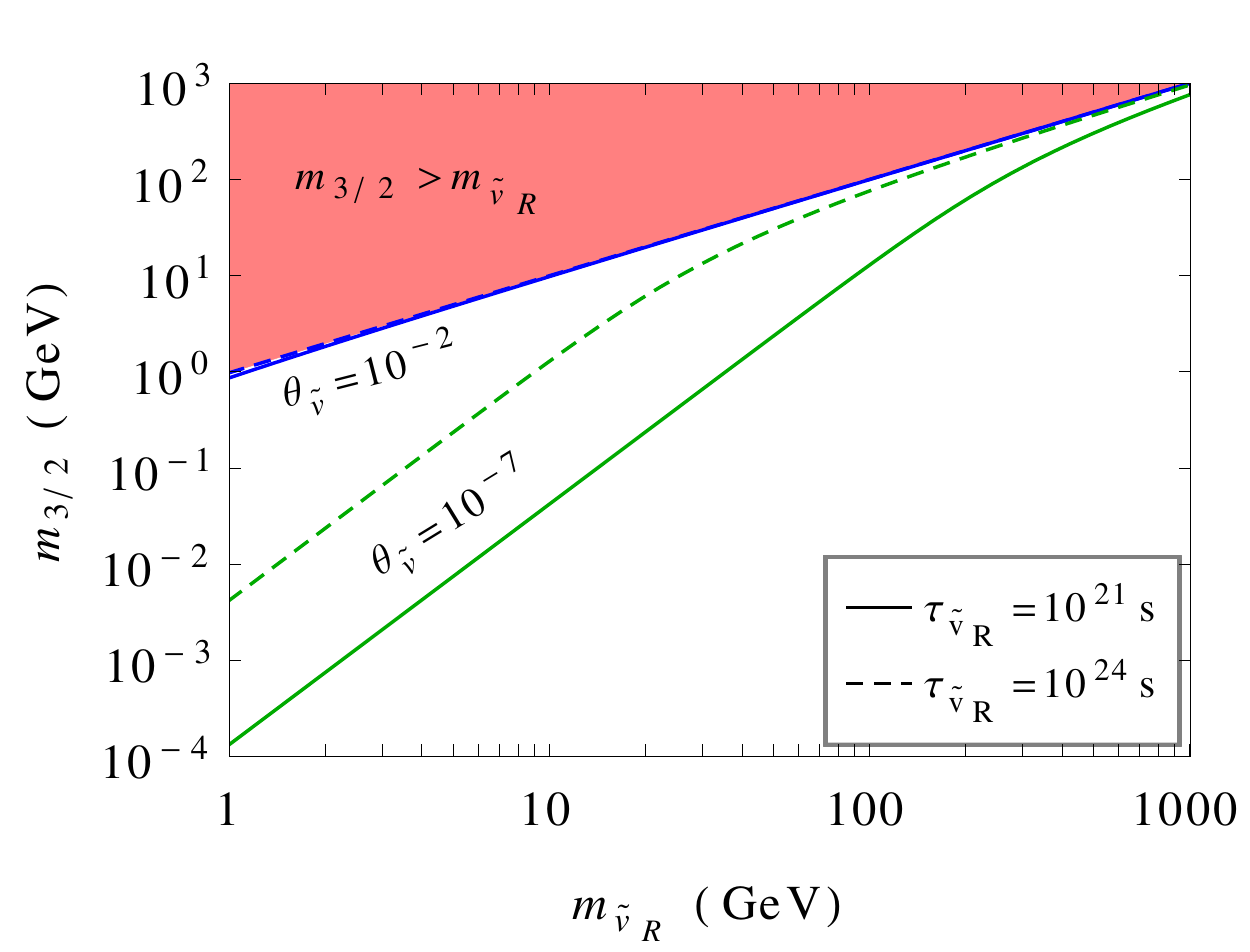,height=7cm} 
    \end{tabular}
    \captions{Gravitino LSP mass needed to get RH sneutrino NLSP with fixed lifetime $\tau_{\tilde{\nu}_R}=10^{21}$ s (solid curves) and $\tau_{\tilde{\nu}_R}=10^{24}$ s (dashed curves). Representative mixing angles between RH and LH sneutrinos are shown (see Sections~\ref{sec:sneutrinosector} and \ref{sec:LHRHmixing}): $\theta_{\tilde{\nu}}=10^{-2}$ (blue) for MSSM+RHN with thermal RH sneutrino DM, i.e. large trilinear neutrino couplings, and $\theta_{\tilde{\nu}}=10^{-7}$ (green) for NMSSM+RHN.
}
    \label{sneutgrav-01}
\end{center}
\end{figure}

To start the analysis Fig.~\ref{sneutgrav-01} shows the expected gravitino mass range predicted by the two models that include RH neutrino superfields discussed in Section~\ref{sec:modelandscan}, for typical mixing angles between LH and RH sneutrinos:
\begin{itemize}
\item $\theta_{\tilde{\nu}}=10^{-2}$ (blue) for MSSM+RHN with viable thermal RH sneutrino DM, i.e. large trilinear neutrino coupling $T_N \sim O(\text{EW})$, and
\item $\theta_{\tilde{\nu}}=10^{-7}$ (green) for the NMSSM+RHN. 
\end{itemize}
For both models we show the gravitino mass required to get RH sneutrino NLSP with lifetime equal to $\tau_{\tilde{\nu}_R}=10^{21}$ s (solid curves) and $\tau_{\tilde{\nu}_R}=10^{24}$ s (dashed curves) given by Eq.~(\ref{sneutgravdecaywidth}). As we will show, these lifetime values are in the ballpark of cutting edge neutrino detectors for energy $O(\text{GeV})$.

We can clearly see that for the MSSM+RHN case, large phase space suppression is needed with the gravitino mass very similar to the RH sneutrino mass to achieve long lifetimes. Also the model is not very sensitive to different lifetimes and therefore it is difficult to explore significant parameter regions with indirect detection experiments. 
On the other hand, for the NMSSM+RHN case we have a mixed scenario with an approximately massless gravitino (but not extremely light) for low RH sneutrino masses, and a region with $m_{3/2}\sim m_{\tilde{\nu}_R}$ for $m_{\tilde{\nu}_R} \sim O(100\text{ GeV})$. This model turns out to be very interesting not only because the gravitino, as well as the RH sneutrino, can be a cold DM candidate and a main component of the current relic density, but also the parameter space of the model is within reach of neutrino telescopes and sensitive enough to probe important regions with current and next generation experiments.

\begin{figure}[t!]
\begin{center}
 \begin{tabular}{ccc}
 \hspace*{-12mm}
 \epsfig{file=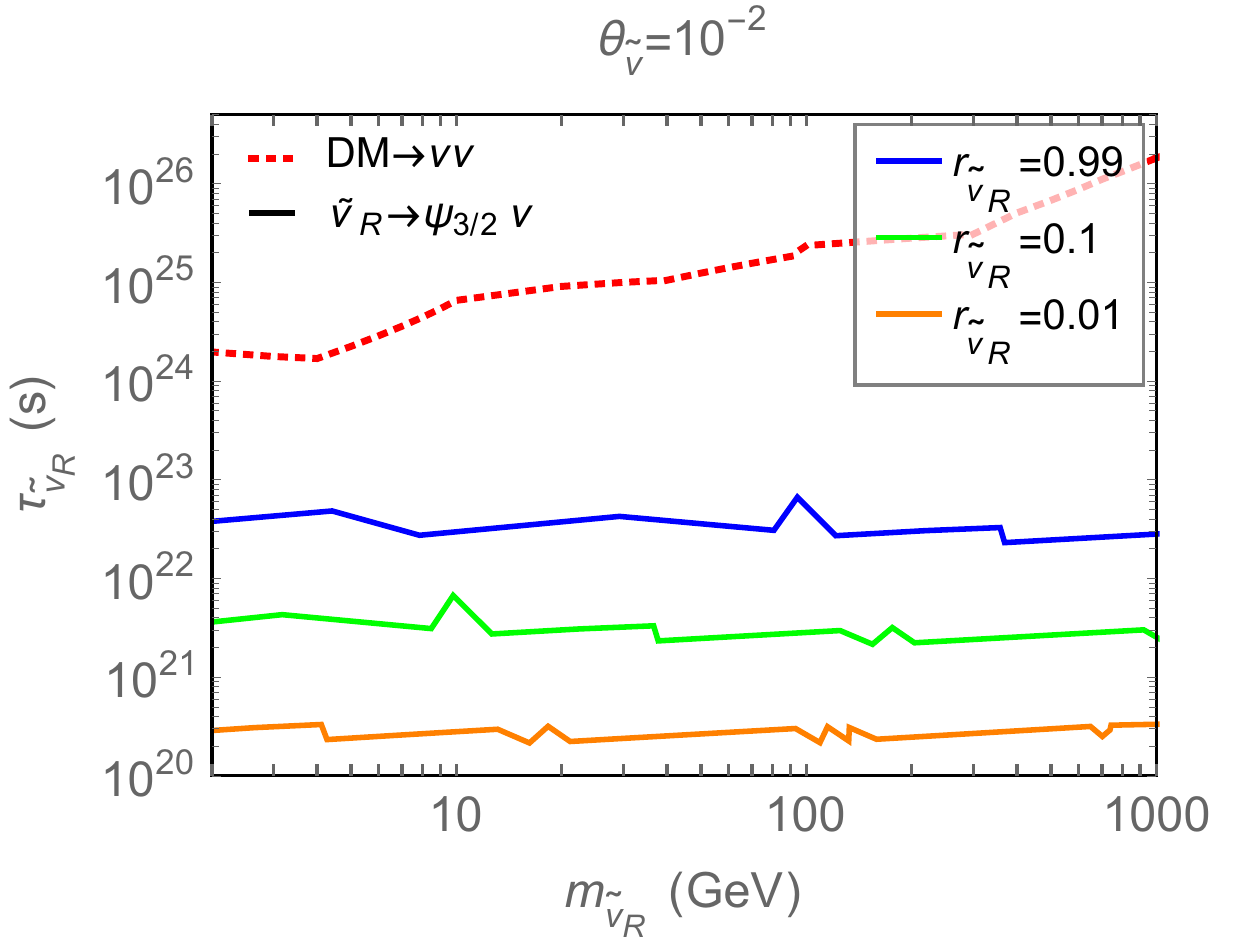,height=7cm} 
 \hspace*{-6mm}
 \epsfig{file=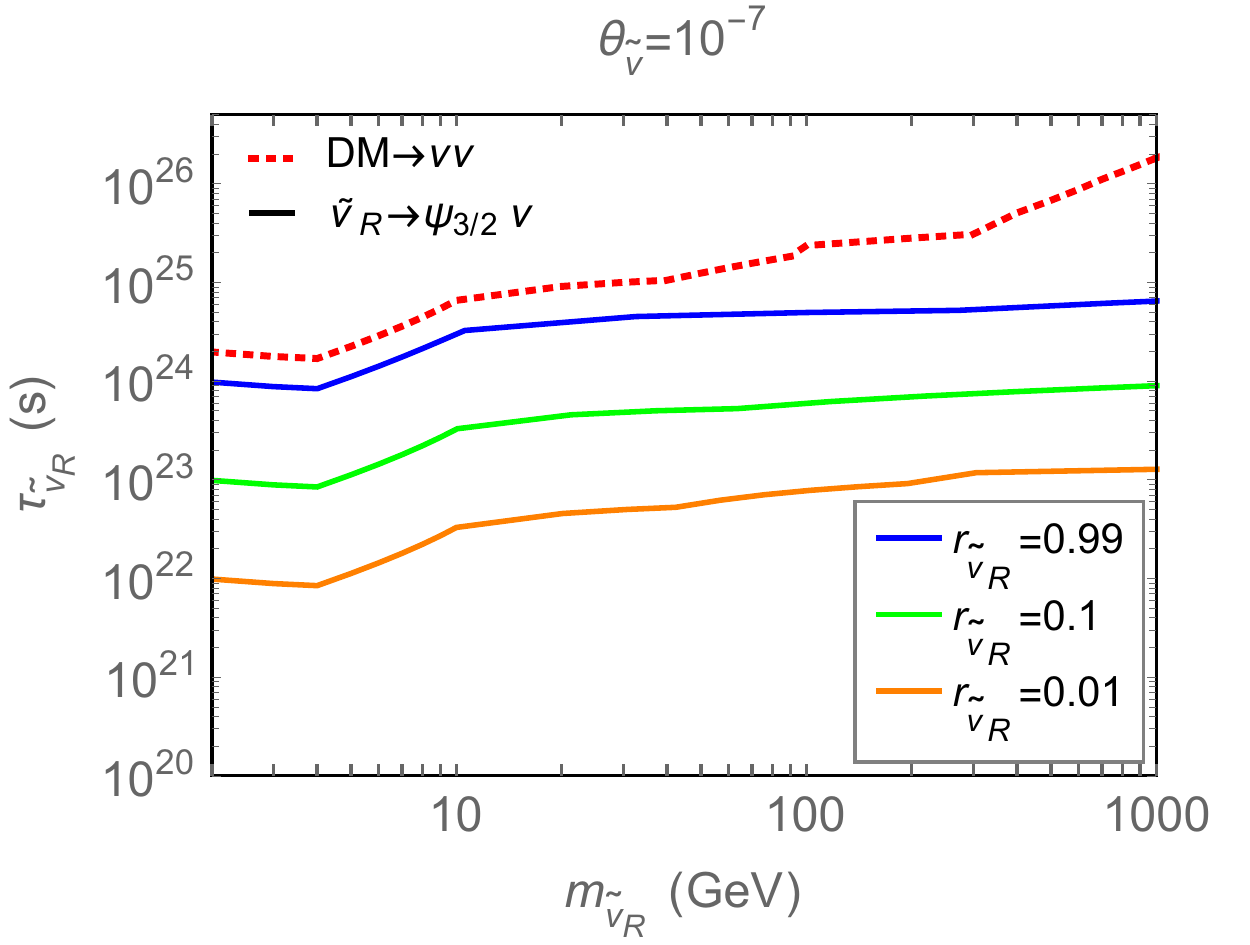,height=7cm}  
    \end{tabular}
    \captions{Current constraints on the sneutrino lifetime given by IceCube~\cite{IceCube:2017rdn}, Antares~\cite{ANTARES:2019svn}, and Super-Kamiokande~\cite{Super-Kamiokande:2020sgt,Olivares-DelCampo:2017feq,Arguelles:2019ouk}. The red dashed curve corresponds to the published constraints considering a DM candidate with $r_{\text{DM}}=1$ decaying to a pair of active neutrinos, with branching ratio equal to 1 (see Appendix~\ref{appen1} for details). The solid curves correspond to the lower limits on sneutrino DM to gravitino plus neutrino with $\theta_{\tilde{\nu}}=10^{-2}$ and $10^{-7}$ for the left and right panel, respectively. Several values of $r_{\tilde{\nu}_R}$ are shown.
}
    \label{sneutgrav-02}
\end{center}
\end{figure}

The current most stringent constraints on neutrino flux in the energy range $\sim O(\text{GeV-TeV})$ are given by IceCube~\cite{IceCube:2017rdn}, Antares~\cite{ANTARES:2019svn}, and Super-Kamiokande~\cite{Super-Kamiokande:2020sgt,Olivares-DelCampo:2017feq,Arguelles:2019ouk}, and shown in Fig.~\ref{sneutgrav-02}. The red dashed curve corresponds to the published lower limit constraints estimated in Appendix~\ref{appen1} by adapting annihilation to decay limits. These constraints consider a sole DM candidate decaying to a pair of active neutrinos (hence $E_{\nu}=m_{DM}/2$) with branching ratio equal to 1. Then, the region below the curve would be excluded.

As discussed in Section~\ref{sec:neutrinosignal}, we have to recast these bounds since we consider a multicomponent scenario with significant gravitino and RH sneutrino relic densities, and only one monochromatic neutrino produced in each decay whose energy depends on the masses of both DM candidates (see Eq.~(\ref{neutrinoenergy})). The modified constraints are also shown in Fig.~\ref{sneutgrav-02} for the models discussed. The new lower limits are parameterized by their typical mixing angles between LH and RH sneutrinos, $\theta_{\tilde{\nu}}=10^{-2}$ and  $10^{-7}$, as labeled on the left and right panels, respectively. We also show for each case several RH sneutrino relic density fractions $r_{\tilde{\nu}_R}$. 


The effects of the multicomponent scenario can be clearly seen in the left panel with $\theta_{\tilde{\nu}}=10^{-2}$. This regime requires $m_{3/2}\sim m_{\tilde{\nu}_R}$, then the bounds presented by the experimental collaborations are modified due to the shift in the neutrino energy. For a fixed $m_{\tilde{\nu}_R}$ value in our multicomponent scenario a monochromatic neutrino with energy $E_{\nu}\simeq m_{\tilde{\nu}_R}-m_{3/2}$ is produced, however the energy of the signal would be $E_{\nu}\simeq m_{DM}/2$ in the massless case. Therefore the limits in our case for $m_{\tilde{\nu}_R}$ turn out to match the lower limit corresponding to a much smaller $m_{\textit{DM}}$ value in the published analysis. 

On the other hand, the right panel of Fig.~\ref{sneutgrav-02} presents the constraints for the NMSSM+RHN with a typical value of $\theta_{\tilde{\nu}}$, where we have a mixture of effects.  For high RH sneutrino masses, $m_{\tilde{\nu}_R}\gtrsim 75$ GeV, we needed $m_{3/2} \sim m_{\tilde{\nu}_R}$ to achieve sneutrinos with lifetimes longer than the age of the Universe (see Fig.~\ref{sneutgrav-01}), and as before, the shift in the neutrino energy is the main effect. For low RH sneutrino masses, the gravitino can be considered massless. In that region the bounds presented by the experimental collaborations are modified taking into account that only one neutrino is produced in each process and that we may have a lower flux for $r_{\tilde{\nu}_R} \leq 1$. It results in weaker constraints, while preserving the same functional form as those published.

\begin{figure}[t!]
\begin{center}
 \begin{tabular}{cc}
 \hspace*{-12mm}
 \epsfig{file=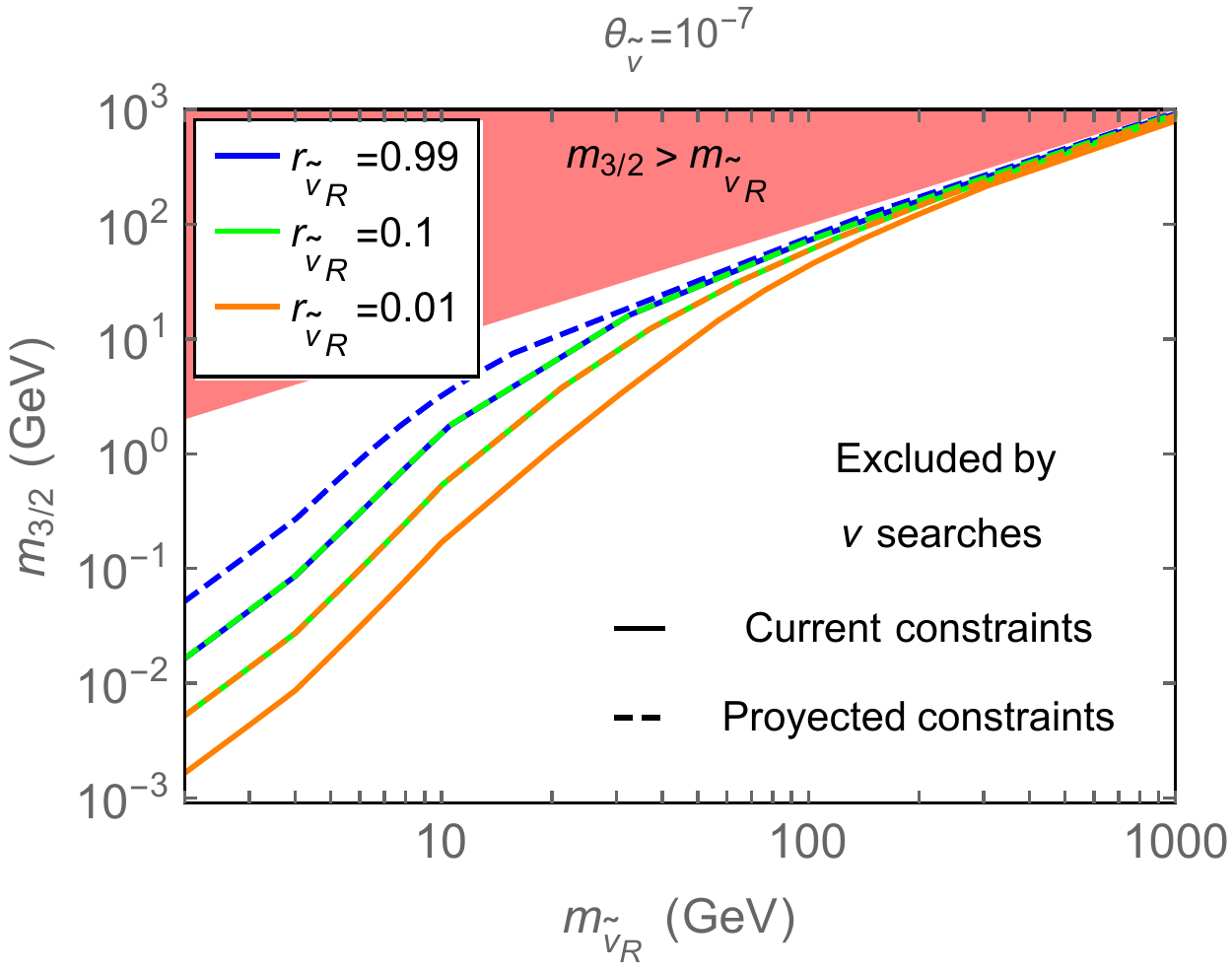,height=7cm} 
       \epsfig{file=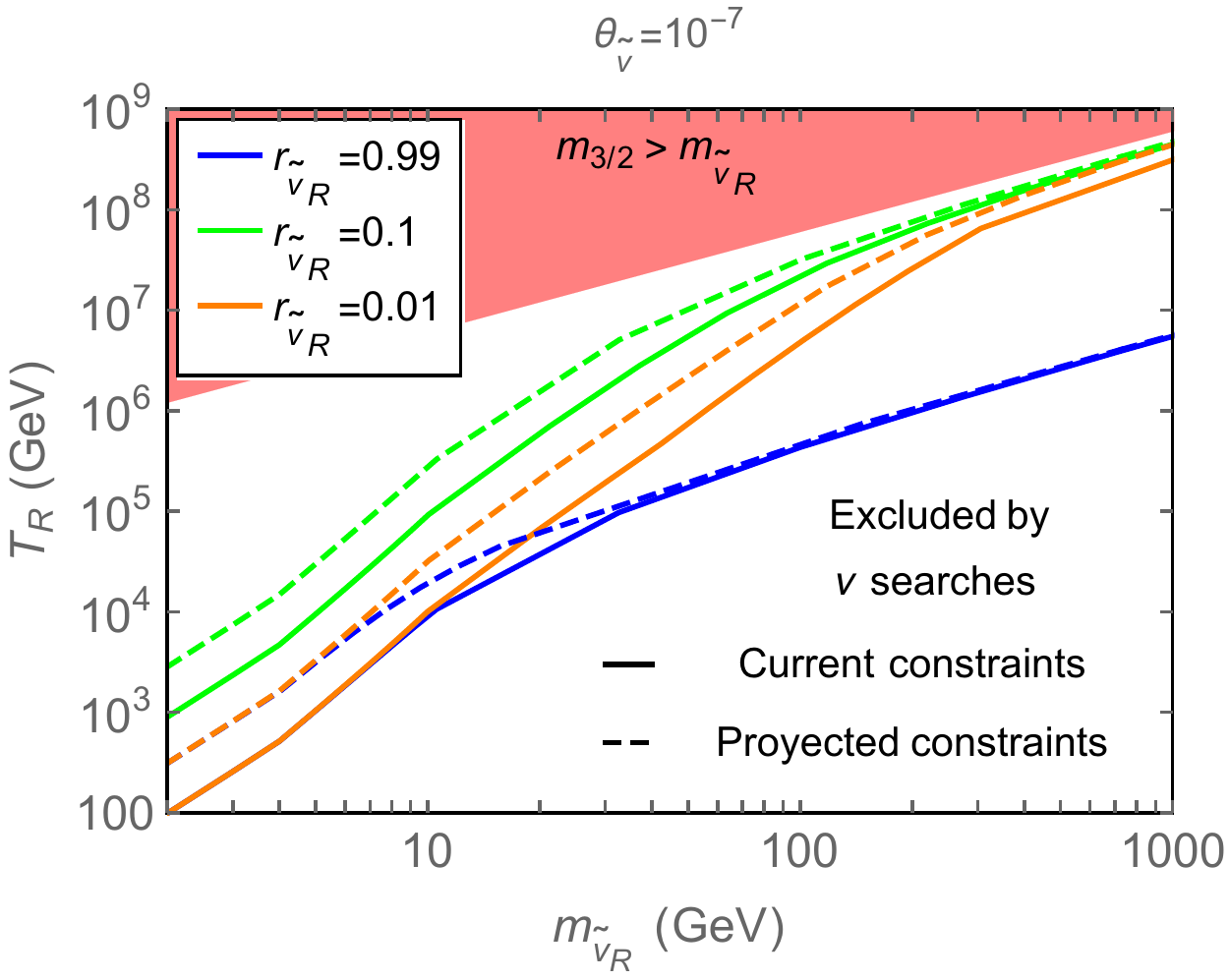,height=7cm}   
    \end{tabular}
    \captions{Allowed parameter regions for the NMSSM+RHN by current and projected $\nu$ searches. For both panels $\theta_{\tilde{\nu}}=10^{-7}$, and several RH sneutrino relic density fractions $r_{\tilde{\nu}_R}$ are depicted as solid curves for current facilities and as dashed curves for future ones. The regions below the curves are excluded. Left panel: gravitino mass lower limit as a function of the RH sneutrino mass, to get an allowed neutrino flux due to RH sneutrino decay. Right panel: reheating temperature lower limit, to satisfy $r_{\tilde{\nu}_R}+r_{3/2}=1$ and the neutrino flux constraint. 
}
    \label{sneutgrav-03}
\end{center}
\end{figure}

Next we will focus on the NMSSM+RHN, since in this model we can have coexisting RH sneutrino NLSP and gravitino LSP, both as cold DM candidates
, and a promising prospect of detection (unlike the MSSM+RHN with thermal DM, where $m_{3/2}\sim m_{\tilde{\nu}_R}$ in the entire parameter space). Then in Fig.~\ref{sneutgrav-03} we show the allowed parameter regions for the NMSSM+RHN. Current constraints (solid curves) are translated as lower limits for the gravitino mass and/or the reheating temperature. We also depict the projected sensitivity of upcoming detectors or updates (dashed curves) taking one order of magnitude improvement with respect to their predecessors, a conservative estimate according to Ref.~\cite{Covi:2010}. We have considered $\theta_{\tilde{\nu}}=10^{-7}$, and depicted several RH sneutrino relic density fractions $r_{\tilde{\nu}_R}$. Important parameter regions are being probed by current and future neutrino facilities.

The left (right) panel of Fig.~\ref{sneutgrav-03} shows the gravitino mass (reheating temperature) lower limit as a function of the RH sneutrino mass, satisfying $r_{\tilde{\nu}_R}+r_{3/2}=1$ and the neutrino flux constraint. The colored region is excluded by overproduction of gravitinos unless $m_{3/2}>m_{\tilde{\nu}_R}$. For low sneutrino masses the most stringent constraints on the right panel correspond to $r_{\tilde{\nu}_R}=0.1$, while for higher sneutrino masses to $r_{\tilde{\nu}_R}=0.01$. This is due to the dependence in $\tau_{\tilde{\nu}_R\text{-eff}}$ and $r_{3/2}$, with $m_{3/2}$ and $r_{\tilde{\nu}_R}$, resulting in $T_R \propto (1-r_{\tilde{\nu}_R}) \, r_{\tilde{\nu}_R}^{1/2}$.

\begin{figure}[t!]
\begin{center}
 \begin{tabular}{cc}
 \hspace*{-12mm}
 \epsfig{file=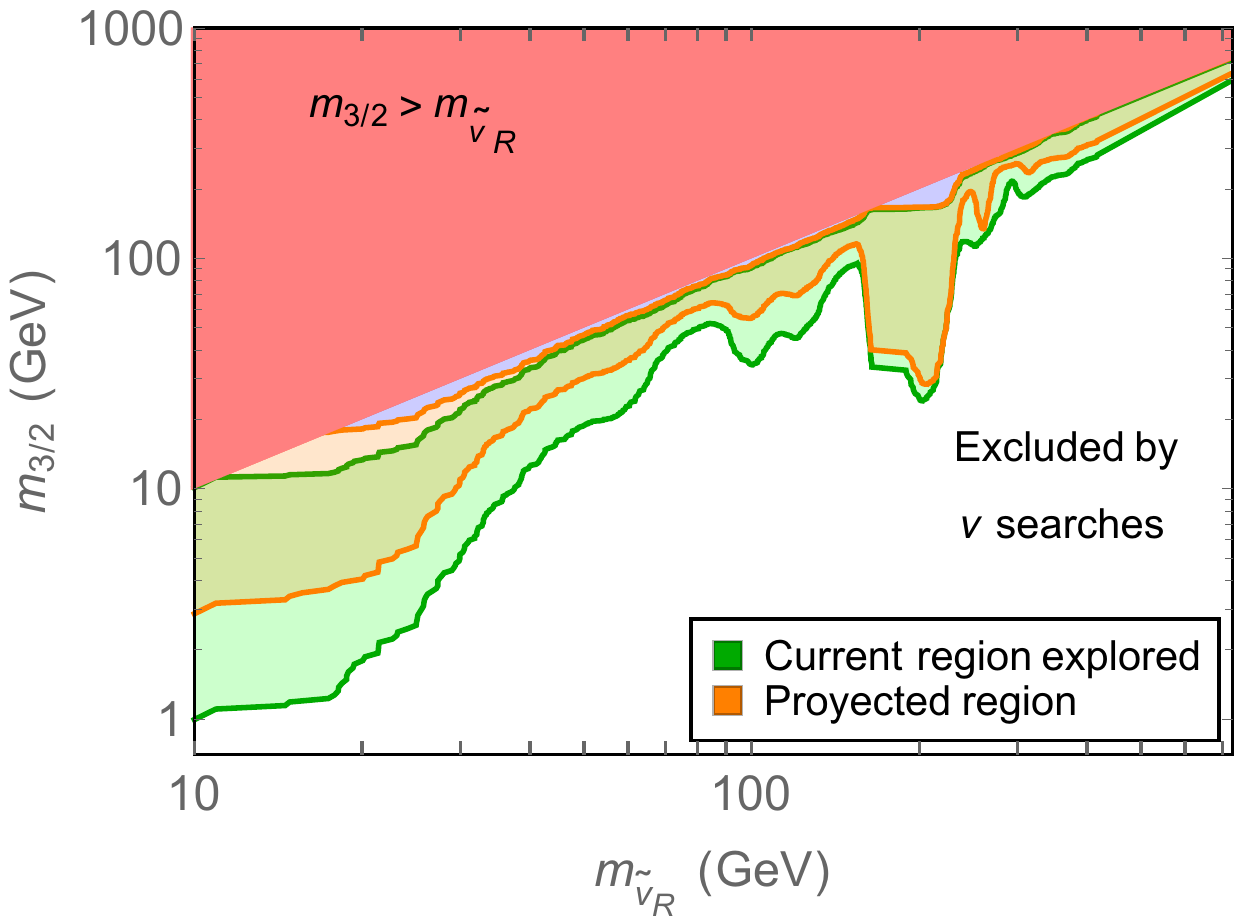,height=6.5cm} 
       \epsfig{file=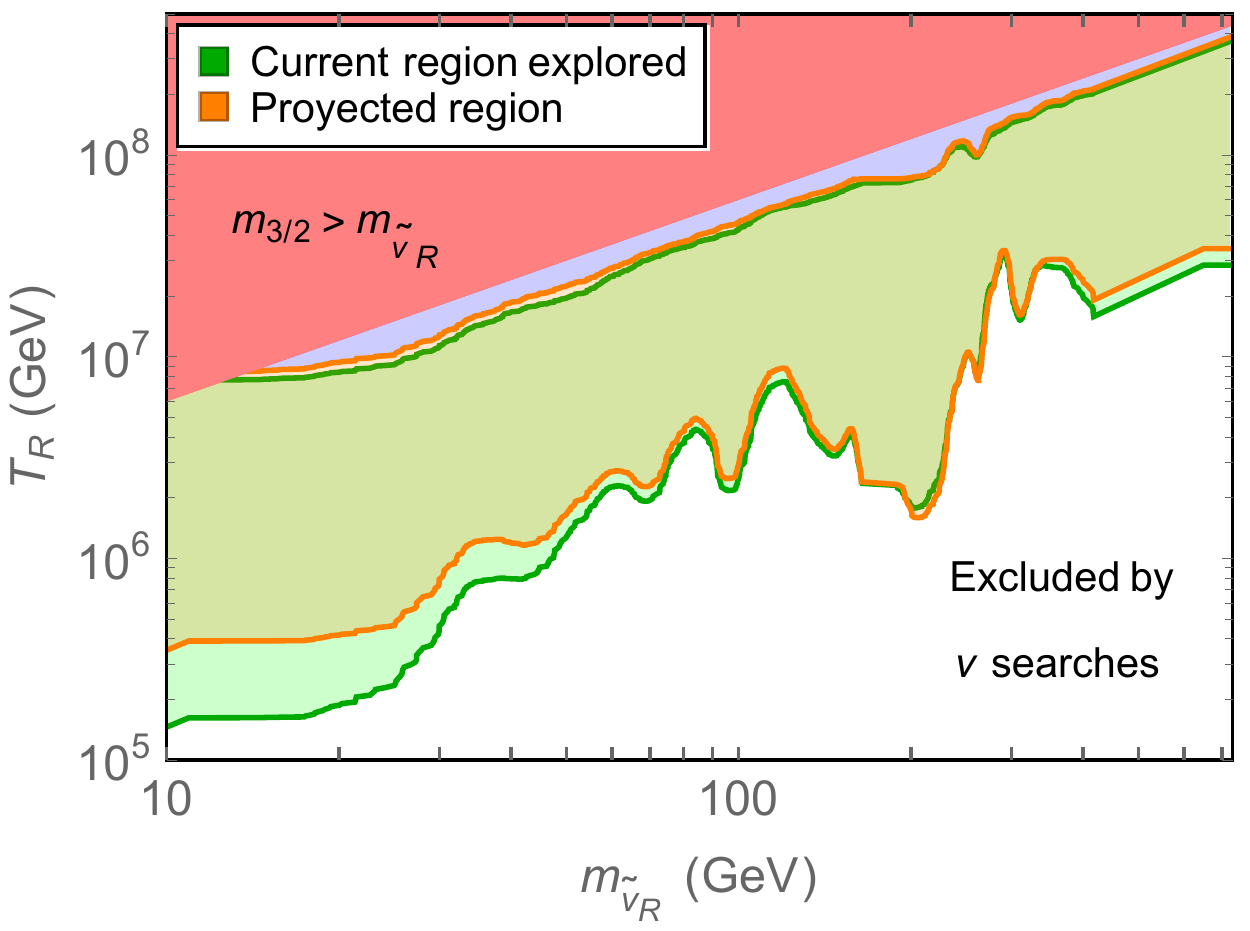,height=6.5cm}   
    \end{tabular}
    \captions{Region currently being explored (green) by neutrino telescopes and region to be probed by upgrades or next generation neutrino experiments (orange), considering the NMSSM+RHN dataset shown in Fig.~\ref{sneutgrav-scan} that obtains an allowed RH sneutrino DM candidate with $r_{\tilde{\nu}_{R}} > 0.01$. Each colored region covers the points obtained with the scans. Below the lower green solid curve no solutions were found that satisfied all the considered experimental constraints, then the white region is excluded by current $\nu$-line searches. The lower orange solid curve would become the projected new lower limit if no signal is detected. The light blue regions above solid upper curves present no numerical solutions and are allowed but the upcoming experiments will not be able to probe them. The right panel was obtained assuming a gluino mass parameter $M_3=3$ TeV in Eq.~(\ref{relicgravitinos}).
}
    \label{sneutgrav-scan-regions}
\end{center}
\end{figure}

Finally, we show our main results in Fig.~\ref{sneutgrav-scan-regions}. We present the allowed parameter space for the coexisting RH sneutrino and gravitino DM scenario in the context of the NMSSM+RHN, using the dataset shown in Fig.~\ref{sneutgrav-scan}. These figures are similar to Fig.~\ref{sneutgrav-03}, where $\theta_{\tilde{\nu}}$ and $r_{\tilde{\nu}}$ were fixed. However, now both are treated as parameters determined by the scan and could take any value, as long as the parameter point satisfies all the experimental constraints considered.

The region shaded in green is currently being explored by neutrino telescopes. The orange region will be probed by upgrades or upcoming neutrino experiments, and will push the explored region almost up to the $m_{3/2} \sim m_{\tilde{\nu}_{R}}$ limit. We would like to highlight that each colored region covers all the points obtained with the NMSSM+RHN scans considering typical parameter values, and that the RH sneutrinos within these solutions satisfy all the usual WIMP constraints. Below the lower green solid curve no solutions were found, therefore the white region is currently excluded by $\nu$-line searches. Assuming a gluino mass parameter $M_3 = 3$ TeV we found the following lower limits for the gravitino mass $m_{3/2} \gtrsim 1-600$ GeV and the reheating temperature $T_R \gtrsim 10^5 - 3 \times 10^7$ GeV, considering $m_{\tilde{\nu}_R} \sim 10-800$ GeV. In the same way, the lower orange solid curve would become the projected new lower limit if no signal is detected, with a bigger improvement (up to a factor of $\sim 3$) for lower masses. Notice that between the explored parameter space and the $m_{3/2} = m_{\tilde{\nu}_{R}}$ line, we show regions in light blue where also no numerical solutions were found during the scan. These regions are allowed but next generation neutrino experiments will not be able to probe them.

\begin{figure}[t!]
\begin{center}
 \begin{tabular}{c}
 \hspace*{-4mm}
 \epsfig{file=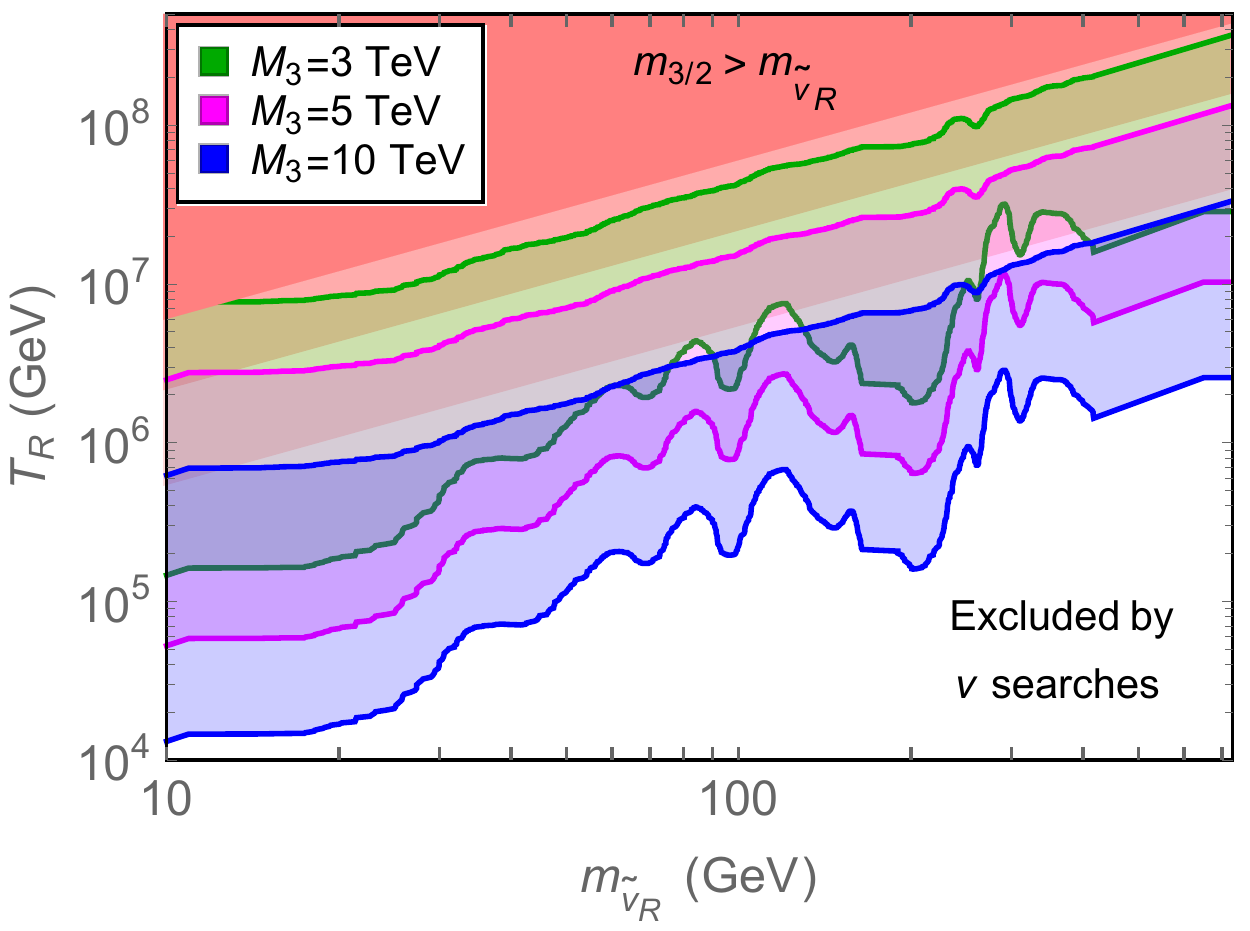,height=7cm} 
    \end{tabular}
    \captions{Effect of the gluino mass parameter, $M_3$, (see Eq.~(\ref{relicgravitinos})) on the $T_R$ versus $m_{\tilde{\nu}_{R}}$ space. Three values are considered $M_3=3, 5,$ and $10$ TeV in green, magenta, and blue, respectively. In this figure each region corresponds to the sensibility that will be probed by the next generation of neutrino telescopes (the $M_3=3$ TeV case is shown on the right panel of Fig.~\ref{sneutgrav-scan-regions}). Notice that the curve denoting the $m_{3/2}>m_{\tilde{\nu}_{R}}$ also depends on $M_3$, and is always close to the upper solid curves of the corresponding colored region.
}
    \label{sneutgrav-TR-constraint-M3}
\end{center}
\end{figure}

Last but not least, in Fig.~\ref{sneutgrav-TR-constraint-M3} we show the effect of the gluino mass parameter $M_3$, (see Eq.~(\ref{relicgravitinos})) on the $T_R$ versus $m_{\tilde{\nu}_{R}}$ space, considering three values $M_3=3, 5,$ and $10$ TeV. Each region corresponds to the points that will be probed by the next generation of neutrino telescopes. In this case, the curve denoting the $m_{3/2}>m_{\tilde{\nu}_{R}}$ also depends on $M_3$, and follows the parameter space to be probed. We can see that for $M_3=10$ TeV the lower limits on $T_R$ are relaxed by one order of magnitude with respect to the $M_3=3$ TeV example. The $m_{3/2} - m_{\tilde{\nu}_{R}}$ space is not affected by $M_3$ since it is only involved in the gravitino relic density computation, and hence the results shown in the left panel of Fig.~\ref{sneutgrav-scan-regions} remain unchanged.

\section{Conclusions}
\label{sec:conclusions}

In this work we have analyzed an scenario with two DM candidates contributing significantly to the relic density of the Universe in the context of $R$-parity conserving SUSY models: RH sneutrinos NLSP and gravitinos LSP. 

The RH sneutrino NLSP turns out to be unstable and decays to a gravitino LSP plus a LH neutrino, through an interaction that involves the mixing between the scalar partners of the LH and RH neutrinos. However, if the mixing parameter is small ($\ll O(10^{-2})$), and the fact that the interaction is suppressed by the Planck mass, the RH sneutrino can have a lifetime longer than the age of the Universe. Therefore, both the RH sneutrino and the gravitino can coexist and be considered as DM candidates. Additionally, the gravitino LSP is essentially decoupled from the rest of the spectrum which allows to compute the RH sneutrino abundance as if it were the LSP, ignoring the gravitino and reintroducing it afterwards.

First, we have studied the viability of this scenario taking into account and comparing well-known $R$-parity conserving extensions of the MSSM and NMSSM that not only provide multiple DM candidates, but can explain massive neutrinos in a simple way by adding RH neutrino superfields. Besides the masses of both DM candidates, a key parameter involved in the NLSP decay is the LH-RH mixing angle in the sneutrino sector, $\theta_{\tilde{\nu}}$. In the MSSM+RHN we have: 
\begin{itemize}
    \item $\theta_{\tilde{\nu}} \sim O(0.01-1)$ for weak-scale trilinear sneutrino couplings, $T_N=A_N Y_N \sim O(\text{EW})$, even with small neutrino Yukawa coupling, resulting in thermal RH sneutrino DM.
\end{itemize}
On the other hand, in the NMSSM+RHN we have estimated
\begin{itemize}
    \item $\theta_{\tilde{\nu}} \sim O(10^{-8}-10^{-6})$, for typical parameter values, i.e. without taking an extremely small neutrino Yukawa coupling nor too large trilinear neutrino couplings.
\end{itemize}
In this model, $Y_N \sim O(10^{-6})$ to obtain the right order for the neutrino masses through a see-saw mechanism, and the RH sneutrino can be thermal DM thanks to extra interaction terms in the Lagrangian between RH neutrino and singlet superfields, absent in MSSM extensions. This $\theta_{\tilde{\nu}}$ range was confirmed employing the results of several scans taking into account the latest collider and astrophysical constraints~\footnote{To facilitate the numerical calculation, we have assumed a diagonal Yukawa matrix, $Y_N^{ii} = 10^{-6}$, then we reproduce not the mixing angles and mass differences of the neutrino sector, but the correct order of its masses. As already mentioned in the main text, this does not change our results, because the main focus of this work is to find viable points with RH sneutrino DM and to estimate the range of $\theta_{\tilde{\nu}}$, since in the model the DM and the neutrino sector are decoupled (for details see Appendix~\ref{appendixneutrino})}.

Interestingly, the decay of the NLSP produces a new signal that is characteristic of these multicomponent models: monochromatic neutrinos in the ballpark of current and planned neutrino telescopes. To probe the proposed signal against published limits, we have recasted the constraints set by Super-Kamiokande, IceCube and Antares since they consider a single DM candidate and an annihilation channel to a pair of neutrinos, i.e. an almost massless final state. On the other hand, in the two-component DM scenario only a fraction of the total DM budget produces neutrinos (RH sneutrinos NLSP) in a decay channel that also involves a massive particle in the final state (gravitino LSP).

Unlike in the MSSM+RHN, where we need to fine-tune the masses to $m_{3/2}\sim m_{\tilde{\nu}_R}$, we have focused on the NMSSM+RHN, since both candidates can be cold DM and important parameter regions can be explored. We have presented our main results in Fig.~\ref{sneutgrav-scan-regions}, where we show the regions that are currently being explored and the ones to be probed in the near future with the next generation of neutrino telescopes. Assuming a gluino mass parameter $M_3 = 3$ TeV we have found the following lower limits for the gravitino mass $m_{3/2} \gtrsim 1-600$ GeV and the reheating temperature $T_R \gtrsim 10^5 - 3 \times 10^7$ GeV, for $m_{\tilde{\nu}_R} \sim 10-800$ GeV. These regions satisfy all current neutrino searches, along with the usual WIMP constraints already considered during the NMSSM+RHN scan for the RH sneutrino DM. If we take $M_3=10$ TeV, then the limits on $T_R$ are relaxed by one order of magnitude.


\section*{Acknowledgments}

The work of DL and AP was supported by the Argentinian CONICET, they also acknowledge the support through PIP 11220170100154CO and PICT 2020-02181. The computational resources used in this work were provided (in part) by the HPC center DIRAC, funded by Instituto de Fisica de Buenos Aires (UBA-CONICET) and part of SNCAD-MinCyT initiative, Argentina. R. RdA is supported by PID2020-113644GB-I00 from the Spanish Ministerio de Ciencia e Innovación.


\appendix

\section{Scan parameters of the NMSSM+RHN}
\label{appen0}

In this appendix we present the parameter ranges employed to find solutions in the context of the NMSSM+RHN with the \texttt{MultiNest}~\cite{multinestcite} algorithm as optimizer.

In the sfermion sector we fix $m^2_{\tilde{e}_i}=m^2_{\tilde{d}_i}=2.25\times 10^{6}$ GeV$^2$ with $i=1,2,3$, $m^2_{\tilde{L}_3} = 2.25\times 10^{6}$ GeV$^2$, and $m^2_{\tilde{N}_i}=m^2_{\tilde{u}_i}=m^2_{\tilde{Q}_i}=2.25\times 10^{6}$ GeV$^2$ with $i=1,2$. These parameters are not especially relevant to our analysis, then their values are taken sufficiently large to be consistent with LHC sparticles searches. Notice that we consider $m^2_{\tilde{L}_1} = m^2_{\tilde{L}_2}$ as an independent parameter in order to reproduce $(g-2)_{\mu}$. If we fix $m^2_{\tilde{L}_2} = 2.25\times 10^{6}$ GeV$^2$ we usually obtain low values of $\delta a_{\mu}$.

The neutrino Yukawa couplings are set $Y_N^{i}=10^{-6}$ when the scan is focused on finding solutions with RH sneutrino LSP. We also consider $T_d^{33}=T_{d_3}=256$ GeV, and $T_e^{33}=T_{e_3}=-98$ GeV taking into account the corresponding Yukawa couplings.  We take vanishing $T_{N}^i=A_N^i \, Y_N^i$ as an approximation due to the small neutrino Yukawa couplings and $A_{_N}^i\sim O(\text{GeV})$. 
To simplify the analysis by obtaining two families of heavy sneutrinos, we set $\lambda_N^i=-0.5$, and $T_{\lambda_N}^{i} = 0$ for $i=1,2$, .

We fix the gluino mass parameter $M_3=3$ TeV to avoid LHC constraints on gluino strong production.

In the Higgs-scalar singlet sector the soft-breaking masses  are related with the vacuum expectation values (VEVs) by the  minimization conditions of the Higgs potential after electroweak symmetry breaking (EWSB). Then, we take as free parameters: the ratio of the Higgs VEVs $\tan \beta\equiv v_u/v_d$, the effective higgsino mass parameter $\mu_{eff}= \lambda \, v_s/ \sqrt{2}$, $\lambda$, $\kappa$,
$T_{\lambda}=A_{\lambda} \, \lambda$ and $T_{\kappa} = A_{\kappa} \, \kappa$.

Therefore, the following variables are considered independent parameters:
\begin{equation}
M_1, \: M_2, \: \tan \beta , \: \mu_{eff}, \: \lambda , \: \kappa , \: \lambda_N^3 , \: T_{\lambda}, \: T_{\kappa}, \: T_{\lambda_N}^3 , \: m^2_{\tilde{L}_2}, \: m^2_{\tilde{N}_3}, \: m^2_{\tilde{u}_3}, \: m^2_{\tilde{Q}_3}, \: T_{u_3},
\label{inputsvariables}
\end{equation}
whose ranges can be seen in the second column labeled as \emph{Scan 1} in Table \ref{scanparameters}. 

To sample the model more efficiently, we performed further scans varying a subset of nine parameters:
\begin{equation}
    M_1, \: M_2, \: \tan \beta, \: \mu_{eff}, \: \lambda,  \: \kappa, \: \lambda_N^3, \: m^2_{\tilde{L}_2}, \: m^2_{\tilde{N}_{3}},
\end{equation}
whose subranges can be seen on the third column of Table~\ref{scanparameters}, labeled as \emph{Scan 2}. We use the \emph{Scan 1} solutions as seeds and denote with a superscript `0' the seed value around which we define our new subrange for the \emph{Scan 2} (or we fix the seed value if the parameter is not considered as an independent variable).

\begin{table}[t]
      \centering
        \hspace{0.9cm}\begin{tabular}{|c|c|c|}
        \hline
        \textbf{Parameter} & \textbf{Scan 1} & \textbf{Scan 2} \\
        \hline
        \hline
         & & \\[-2ex]
            $M_1$ & (20, 3000) GeV & (40, 1500) GeV \\[1ex]
            $M_2$ & (20, 3000) GeV & (100, 1500) GeV \\[1ex]
            $\mu_{eff}$ & (100, 5000) GeV & $\mu_{eff}^0\pm$10\% \\[1ex]
            $\tan \beta$ & (2, 50) & ($\tan \beta^0-10\%$, 50) \\[1ex]
            $\lambda$ & (0.001, 0.8) & $\lambda^0\pm$10\% \\[1ex]
            $\kappa$ & (0.001, 0.8) & $\kappa^0\pm$10\% \\[1ex]
            $\lambda_N^3$ & (-0.4, -0.001) & $\lambda_N^{3^0}\pm$10\% \\[1ex]
            $T_{\lambda}$ & (0.001, 600) GeV & $T_{\lambda}^0$ \\[1ex]
            $T_{\kappa}$ & (-30, -0.001) GeV & $T_{\kappa}^0$ \\[1ex]
            $T_{\lambda_N}^3$ & (-1100, -0.001) & $T_{\lambda_N}^{3^0}$ \\[1ex]
            $m^2_{\tilde{N_3}}$ & ($10$, $2.5\times10^{6}$) GeV$^2$ & $m^2_{\tilde{N_3}^0}\pm$10\% \\[1ex]
            $m^2_{\tilde{u}_3}$ & ($2.5\times10^{5}$, $4\times10^{6}$) GeV$^2$ & $m^2_{{{\tilde{u}_3}}^0}$ \\[1ex]
            $m^2_{\tilde{Q}_3}$ & ($2.5\times10^{5}$, $4\times10^{6}$) GeV$^2$ & $m^2_{{\tilde{Q}_3}^0}$ \\[1ex]
            $m^2_{\tilde{L}_1}$ = $m^2_{\tilde{L}_2}$ & $2.25\times10^{6}$ GeV$^2$ & (($\mu_{eff}^0-50 \text{GeV}$)$^2$, $1\times10^{6}$) GeV$^2$ \\[1ex]
            $T_{u_3}$ & (700, 10000) GeV & $T_{u_3}^0$ \\[1ex]
        \hline
        \end{tabular}
   \caption{Range of the parameters used in our scan. In \emph{Scan 2}, the superscript `0' refers to the seed value of the corresponding parameter found in \emph{Scan 1}.}
  \label{scanparameters}
\end{table}

\subsection{About numerical solutions for neutrino physics}
\label{appendixneutrino}

We would like to highlight that, to ease the computation, we have fixed the neutrino Yukawa couplings to $Y_N^{ii}=10^{-6}$ with $i=1,...,3$ and zero otherwise. With this set-up we reduce the already large number of free parameters of the model but we do not reproduce the measured neutrino physics, although with the considered fixed values we obtain the correct order of magnitude for the neutrino masses (see Eq.~(\ref{neutrinomassNMSSMplusRHN})).

The main focus of the scan is to find solutions with RH sneutrino as a good DM candidate. Importantly, we can consider the RH sneutrino sector decoupled from the active neutrino one, since the main channels to obtain an allowed relic density involve neutralinos, resonances, or interactions with the Higgs-scalar sector. In particular, for low RH sneutrino masses, the annihilation with pseudoscalars involves the terms $\lambda_N N N S$ and $\lambda S H H$ of the superpotential. Moreover the RH sneutrino mass is independent on $Y_N$ assuming a small LH-RH mixing angle (see Eq.~\ref{RHsneutrinomasses}). Regarding the experimental constraints, the limits on $(g-2)_{\mu}$, Higgs sector and collider searches tested in \texttt{SModelS} are not sensitive to neutrino physics. Also, the RH sneutrino scattering with a nucleon occurs via t-channel exchange of a neutral CP-even Higgs boson, involving the terms $\lambda_N N N S$, $\lambda S H H$, $\kappa S^3$, and $A_{\lambda_N} \lambda_N N N S$.  (for further details about the viability of the RH sneutrino as DM candidate see Refs.~\cite{Kim:2021suj,Lopez-Fogliani:2021qpq}).

For the multicomponent DM scenario considered in this work we also need RH neutrinos heavier than RH sneutrinos to avoid a NLSP with short lifetimes. We have checked that this is satisfied in the NMSSM+RHN scans with $Y_N^{ii}=10^{-6}$. Also, notice that the RH neutrino masses will not be modified significantly changing the neutrino Yukawa matrix since they are dominated by the Majorana mass term $M_N \sim \lambda_N v_s$.

The LH-RH sneutrino mixing, $\theta_{\tilde{\nu}}$, involved in the RH sneutrino decay do depend on the neutrino Yukawa value. Nevertheless, in this work we focus on the typical range for this mixing parameter in the NMSSM+RHN, that differs several orders of magnitud with the one compatible with the MSSM+RHN. If we modify $Y_N$ around its fixed value then $\theta_{\tilde{\nu}}$ with also change but the overall range will remain the same.

Therefore, for each solution with a RH sneutrino DM, one could fix all the parameters to their found values and perform an extra scan over $Y_N^{ij}$ to satisfy the constraints on the mixing angles and mass differences of the neutrino sector. Since we are already considering a parameter space that reproduces the right order of magnitude of the neutrino masses, with the extra scan we expect to find a non-diagonal Yukawa matrix with entries $\sim O(10^{-7}-10^{-6})$. Due to the nature of the neutrino sector, this is a very computationally expensive procedure that would be needed for each point, but the conclusion of this manuscript would not change.

As an example, in Table~\ref{BRneutrinophysics} we show the parameters that satisfy the neutrino mixing angles, mass differences~\cite{Gonzalez-Garcia:2021dve} and sum over their masses~\cite{Planck:2018vyg} for a particular solution with RH sneutrino DM, after a dedicated scan where we fix every parameter but $Y_N^{ij}$.

\begin{table}
\begin{footnotesize}
\begin{center}
\begin{tabular}{|c|c|c|c|c|c|}
        \hline
        \multicolumn{6}{|l|}{\textbf{Parameters}} \\
        \hline
        \hline
            $M_1$ & 390.2 GeV & $\mu_{eff}$ & 140.1 GeV & $m^2_{\tilde{N}_3}$ & 2492 GeV$^2$ \\
            $M_2$ & 1483 GeV & $\lambda$ & 0.070 & $m^2_{\tilde{u}_3}$ & 1.8$\times 10^{6}$ GeV$^2$ \\
            $\tan \beta$ & 15.4 & $\kappa$ & 0.042 & $m^2_{\tilde{Q}_3}$ & 1.6$\times 10^{6}$ GeV$^2$ \\
            $\lambda_N^3$ & -0.161 & $T_{\lambda}$ & 189.6 GeV & $m^2_{\tilde{L}_2}$ & 7434 GeV$^2$ \\
            $T_{\lambda_N}^3$ & -87.0 GeV & $T_{\kappa}$ & -0.020 GeV & $T_{u_3}$ & 1854 GeV \\
        \hline
        \multicolumn{6}{|l|}{\textit{Neutrino sector}} \\
        \hline
            $Y_N^{11}$ & $4.23\times 10^{-7}$ & $Y_N^{12}$ & $4.53\times 10^{-8}$ & $Y_N^{13}$ & $1.38\times 10^{-7}$ \\
            $Y_N^{21}$ & $4.60\times 10^{-7}$ & $Y_N^{22}$ & $1.31\times 10^{-6}$ & $Y_N^{23}$ & $2.36\times 10^{-8}$ \\
            $Y_N^{31}$ & $9.48\times 10^{-7}$ & $Y_N^{32}$ & $7.06\times 10^{-7}$ & $Y_N^{33}$ & $3.19\times 10^{-8}$ \\
        \hline
        \multicolumn{6}{c}{} \\[-1ex]
        \hline
        \multicolumn{6}{|l|}{\textbf{Spectrum}} \\
        \hline
        \hline
            $m_{\tilde{\nu}_{R}}$ & 114.5 GeV & $m_{h_1}$ & 123.5 GeV & $m_{\tilde{\nu}_{L}}$ & 150.6 GeV \\
            $m_{\chi_1^0}$ & 134.7 GeV & $m_{h_2}$ & 273.8 GeV & $m_{\tilde{\mu}_{L}}$ & 169.7 GeV \\
             & \textit{(Higgsino)} & & \textit{(singlet)} & $m_{\nu_{R}}$ & 642.8 GeV \\
            $m_{\chi_2^0}$ & 146.3 GeV & $m_{A_1}$ & 20.8 GeV & & \\
            $m_{\chi_1^{\pm}}$ & 142.5 GeV & & \textit{(singlet)} & & \\
        \hline
        \multicolumn{6}{|l|}{\textit{Neutrino sector}} \\
        \hline
            $\sum_i^3 m_{\nu_i}$ & 0.051 & $\Delta m^2_{21}$ & $7.13\times 10^{-5}$ eV$^2$ & $\Delta m^2_{31}$ & $2.55\times 10^{-3}$ eV$^2$ \\
            $\sin^2\theta_{12}$ & 0.271 & $\sin^2\theta_{13}$ & 0.0241 & $\sin^2\theta_{23}$ & 0.590 \\
        \hline
        
        \hline
        \multicolumn{6}{c}{} \\[-1ex]
        \hline
            $\Omega_{\tilde{\nu}_{R}}h^2$ & 0.110 & $\delta$a$_{\mu}$ & 1.427$\times 10^{-9}$ & $\theta_{\tilde{\nu}}$ & 6.53$\times 10^{-8}$\\
        \hline
\end{tabular}
\end{center}
\end{footnotesize}
   \caption{Benchmark point that satisfies the neutrino physics~\cite{Planck:2018vyg,Gonzalez-Garcia:2021dve} after performing an extra scan fixing every parameter except $Y_N^{ij}$ for a particular solution with RH sneutrino DM. $\delta$a$_{\mu}$ is the $(g-2)_{\mu}/2$ difference between the experimental measurement and theoretical calculation.}
  \label{BRneutrinophysics}
\end{table}

\section{Constraints on DM decay to neutrinos from annihilation limits}
\label{appen1}

\begin{figure}[t!]
\begin{center}
 \begin{tabular}{cc}
 \hspace*{-12mm}
 \epsfig{file=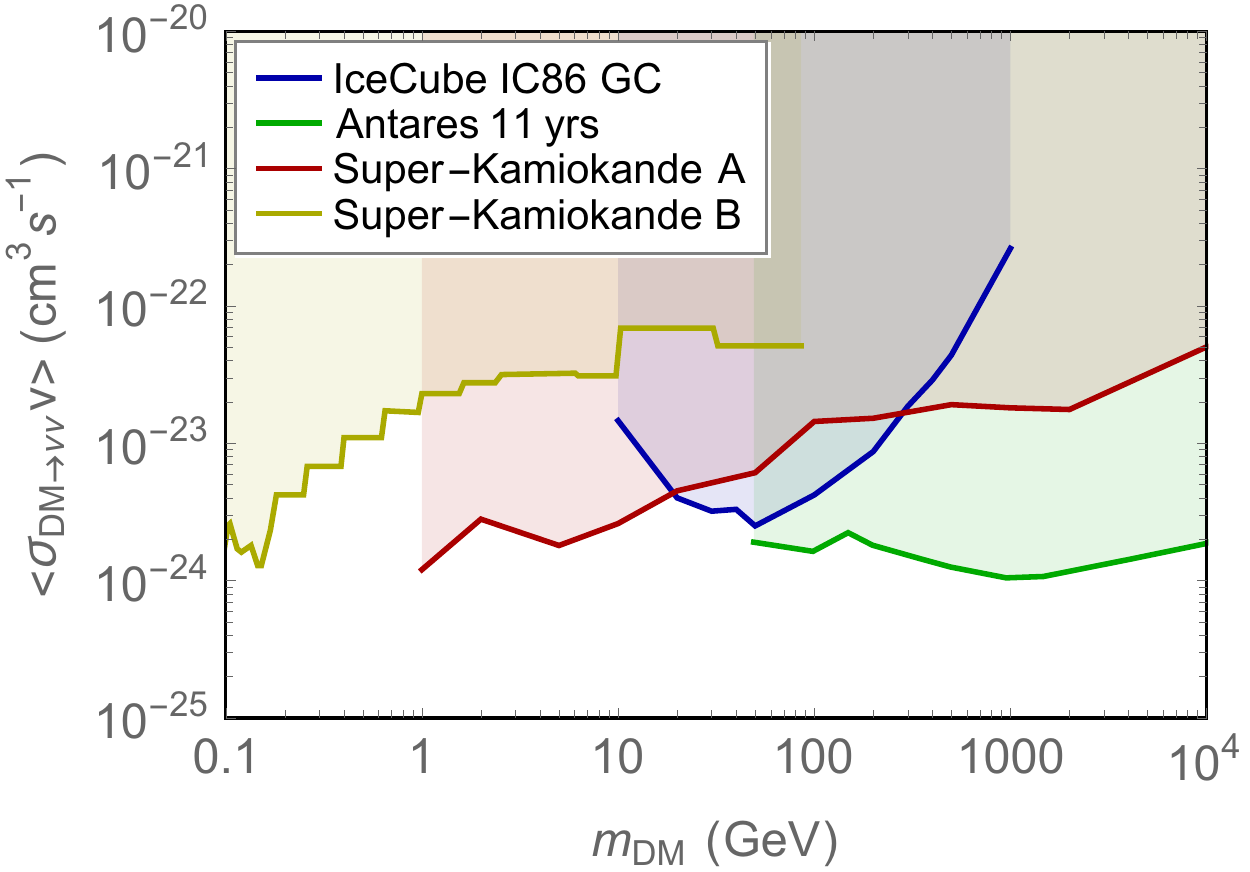,height=6.3cm} 
       \epsfig{file=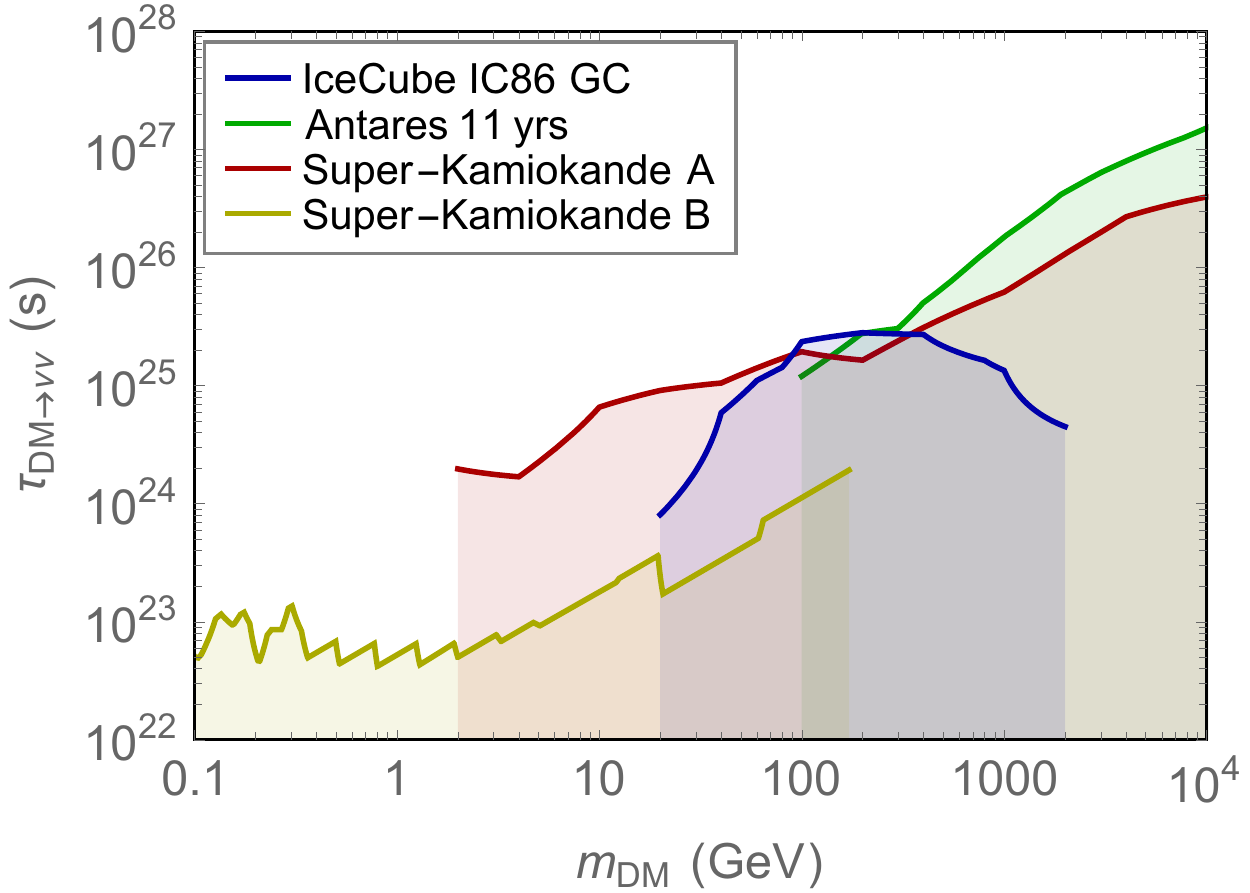,height=6.3cm}   
    \end{tabular}
    \captions{Left panel: upper limits at 90$\%$ C.L. on the DM self-annihilation cross section to a pair of neutrinos, published by IceCube~\cite{IceCube:2017rdn}, Antares~\cite{ANTARES:2019svn}, Super-Kamiokande collaboration~\cite{Super-Kamiokande:2020sgt} (denoted with an `A') and two independent analysis of Super-Kamiokande data~\cite{Olivares-DelCampo:2017feq,Arguelles:2019ouk} (denoted with a `B'). The regions above the lines are excluded. Right panel: lower limits on the DM lifetime considering a decay to a pair of neutrinos calculated from the annihilation cross section upper limits. The regions below the lines are excluded. For both panels, DM halo profile parameters and ROI can be found in Table~\ref{DMprofilesTable}.
}
    \label{constraintsExperiments}
\end{center}
\end{figure}

The experimental collaborations usually present limits on DM considering annihilation processes with neutrinos in the final state. In the left panel of Fig.~\ref{constraintsExperiments} we present the current constraints on DM annihilation to a pair of neutrinos published by IceCube~\cite{IceCube:2017rdn}, Antares~\cite{ANTARES:2019svn}, Super-Kamiokande~\cite{Super-Kamiokande:2020sgt}, denoted with an `A' in the figure, and two independent analysis of Super-Kamiokande data~\cite{Olivares-DelCampo:2017feq,Arguelles:2019ouk} denoted with a `B'. However, to study our scenario with decaying DM we must convert these limits on DM annihilation. In this appendix we outline the procedure we apply to get the constraints.

The differential neutrino flux from DM annihilation or decay in the Galactic halo can be written as
\begin{equation}
  \frac{d\Phi_{\nu}^{\text{halo}}}{dEd\Omega}=\frac{r_{\odot}\, \Gamma}{4\,\pi} \, \left( \frac{\rho_{\odot}}{m_{\textit{DM}}} \right)^{\alpha}\,\frac{dN^{\text{total}}_{\nu}}{dE} \,\frac{1}{\Delta\Omega} \, \int_{\Delta\Omega}\!\!\cos
  b\,db\,d\ell\int_0^{\infty}\!\! \frac{ds}{r_{\odot}}\,\left(\frac{\rho_{\text{halo}}(r(s,\,b,\,\ell))}{\rho_{\odot}}\right)^{\alpha},
  \label{eq:annFlux}
\end{equation}
where $\alpha=1$ ($\alpha=2$) corresponds to DM decay (annihilation). The definition of several parameters can be found in the discussion of Eq.~(\ref{eq:decayFlux}), where the differential flux for the case of DM decay is shown. We would like to highlight three key elements in Eq.~(\ref{eq:annFlux}):
\begin{itemize}

\item The DM interaction rate denoted by $\Gamma$. For DM decay, $\Gamma = 1 / \tau_{DM}$ whereas 
for DM annihilation, $\Gamma = a \, \langle \sigma v \rangle$, where $\langle \sigma v \rangle$ is the velocity-averaged annihilation cross section, and $a=1/2 \, (1/4)$ if the DM candidate is (is not) self conjugated.

\item The neutrino energy spectrum denoted by $\frac{dN^{\text{total}}_{\nu}}{dE} = N_{\nu}^{(\alpha)} \, \delta (E- E_{\nu})$, where $N_{\nu}^{(\alpha)}$ is the number of neutrinos produced in a single process. For DM decay (annihilation), $E_{\nu} = m_{DM}/2$ ($E_{\nu} = m_{DM}$) and we denote $N_{\nu}^{(1)} = N_{\nu}^{(dec)}$ ($N_{\nu}^{(2)} = N_{\nu}^{(ann)}$).

\item The astrophysical part, called D (J)-factor for DM decay (annihilation). This is represented by the integral over $\rho_{\text{halo}}$ raised to the power $\alpha$ along the line of sight {divided by $\Delta\Omega$}. Here $\rho_{\odot}$ denotes the DM density at the location of the Sun $r_{\odot}$, and both parameters are included to make the D- and J-factors dimensionless. 

\end{itemize}

The constraints from neutrino searches considering DM decay (annihilation) are usually presented as lower limits (upper limits) to the particle lifetime (velocity-averaged annihilation cross section) as a function of the DM mass. Also, the bounds are quoted for observations over a specific ROI and DM halo density profile, i.e. a particular D-factor (J-factor).

It is important to note that the signal that can be measured is the same for both processes, decay and annihilation, namely a monochromatic neutrino line if both processes produces a pair of neutrinos. The only difference is that a spectrum with the same energy corresponds to different DM masses, as pointed out in the second bullet point. Then, lifetime lower limits and cross-section upper limits can be obtained from the spectral line flux upper limits~\footnote{The same principle was shown by the Fermi collaboration explicitly in Ref.~\cite{Fermi-LAT:2012ugx} considering $\gamma$-rays.}.
Taking into account all the aforementioned, we can relate the lower limits on $\tau_{DM}$ for DM decay to neutrinos to the upper limits on $\langle \sigma v \rangle$ for DM annihilation to neutrinos (or vice-versa), for the same ROI, as
\bea
\tau_{DM} [m_{DM}] \rightarrow \frac{1}{4 \, a \, \rho_{\odot}} \, \frac{m_{DM}}{\langle \sigma v \rangle [m_{DM}/2]} \, \frac{N_{\nu}^{(dec)}}{N_{\nu}^{(ann)}} \, \frac{\text{D-factor}}{\text{J-factor}},
\label{fractionaxino}
\eea
where the notation $\tau_{DM} [m_{DM}]$ and $\langle \sigma v \rangle [m_{DM}/2]$ implies that the bound on the lifetime corresponding to a DM particle with mass $m_{DM}$ is related to the bound on the annihilation cross section corresponding to a DM particle with mass $m_{DM}/2$.

\begin{table}
\begin{center}
\begin{tabular}{ cccccccc } 
 \hline
  & ROI & D-factor & J-factor \\ 
  \hline
   \hline
 IceCube IC86 GC & RA=[0,2$\pi$], Dec=[-1,1] & 2.18 & 3.94 \\ 
 Antares 11 yrs & 30$^\text{o}$ around the Galactic Center & 6.38 & 29.8 \\  
 Super-Kamiokande A & Full sky & 2.12 & 2.99 \\ 
 Super-Kamiokande B & Full sky & 2.11 & 4.59 \\ 
 \hline
\end{tabular}
 \caption{\label{DMprofilesTable} Values of the J- and D-factors for a NFW profile and the corresponding ROI considered by IceCube~\cite{IceCube:2017rdn}, Antares~\cite{ANTARES:2019svn}, Super-Kamiokande A~\cite{Super-Kamiokande:2020sgt} (from a paper published by the collaboration), and Super-Kamiokande B~\cite{Olivares-DelCampo:2017feq,Arguelles:2019ouk} (independent analyzes). IceCube and Antares take $\gamma = 1$, $r_s=16.1$ kpc, $\rho_s=0.533$ GeV/cm$^3$, $r_{\odot}=8.08$ kpc, and $\rho_{\odot}=0.471$ GeV/cm$^3$. Super-Kamiokande A uses $\gamma = 1$, $r_s=20$ kpc, $\rho_s=0.259$ GeV/cm$^3$, $r_{\odot}=8.5$ kpc, and $\rho_{\odot}=0.3$ GeV/cm$^3$. Super-Kamiokande B $\gamma = 1.2$, $r_s=20$ kpc, $\rho_s=0.251$ GeV/cm$^3$, $r_{\odot}=8.127$ kpc, and $\rho_{\odot}=0.4$ GeV/cm$^3$. $RA$ denotes right ascension and $Dec$ declination.}
\end{center}
\end{table}

In the right panel of Fig.~\ref{constraintsExperiments} we show the constraints for DM decay to a pair of neutrinos ($N_{\gamma}^{(dec)} = 2$), considering the upper limits on $\langle \sigma v \rangle$ to a pair of neutrinos ($N_{\gamma}^{(ann)}=2$) published by IceCube~\cite{IceCube:2017rdn}, Antares~\cite{ANTARES:2019svn}, and Super-Kamiokande~\cite{Olivares-DelCampo:2017feq,Arguelles:2019ouk,Super-Kamiokande:2020sgt}. In every case $a=1/2$ and a generalized NFW~\cite{Navarro:1995iw} DM density profile was considered:
\begin{equation}
\rho_{\text{NFW}}(r) =  \frac{\rho_{s}}{\left(\frac{r}{r_{s}}\right)^{\gamma}\left(1+\frac{r}{r_{s}}\right)^{3-\gamma}}, \\
\label{profiles}
\end{equation}
where $r_s$ and $\rho_s$ represent typical scale radius and scale density, and $\gamma$ parameterizes the generalized profile. Notice that IceCube, Antares and the Super-Kamiokande collaborations take $\gamma = 1$, the standard NFW profile, while the independent analyzes of Super-Kamiokande $\gamma = 1.2$. Moreover, each study adopts different profile parameters $r_s$ and $\rho_s$, and considers a different ROI, whose values are presented in Table~\ref{DMprofilesTable} and its caption. To estimate the lower limits on $\tau_{DM}$ for DM decay in every case we use the J- and D-factors calculated (also shown in Table~\ref{DMprofilesTable}) with the same profile parameter values and the same field of view used in the corresponding $\langle \sigma v \rangle$ determination. We have also calculated the lower limits for DM decay to a pair of neutrinos from two other IceCube publications, Refs.~\cite{IceCube:2015rnn,IceCube:2016oqp}, where a different field of view with respect to Ref~\cite{IceCube:2017rdn} is taken into account. However, they impose weaker constraints than the ones presented in this section.

Finally, we would like to point out that we have followed a conservative approach to compute the lifetime lower limits, since we recast results optimized for annihilation. A dedicated analysis is needed to optimize the signal-to-background ratio, for example determining and using a more efficient ROI for decay processes. We encourage the experimental collaborations to perform these useful studies.


\bibliographystyle{utphys}
\bibliography{munussm}

\end{document}